\documentclass[preprint]{aastex}
\pdfoutput=1 
\newcommand{\um}{\ensuremath{\mu}m} 
\newcommand{\hh}{\mbox{H$_2$}}

\usepackage{fp}

\newcommand{\Sources}{3557}

\newcommand{\Singles}{2935}
\FPsub{\Doubles}{\Sources}{\Singles}
\FPround{\Doubles}{\Doubles}{0}
\FPdiv{\DoubleClumps}{\Doubles}{2}
\FPround{\DoubleClumps}{\DoubleClumps}{0}

\newcommand{\Positions}{3246}
\newcommand{\Classified}{2573}
\FPsub{\Uncertain}{\Positions}{\Classified}
\FPround{\Uncertain}{\Uncertain}{0}

\newcommand{\WithFit}{3218}
\FPsub{\NoFit}{\Positions}{\WithFit}
\FPround{\NoFit}{\NoFit}{0}

\newcommand{\WithFitQuiescents}{616}
\newcommand{\WithFitProtostellars}{749}
\newcommand{\WithFitHiiregions}{844}
\newcommand{\WithFitPDRs}{343}
\newcommand{\WithFitUncertain}{666}

\newcommand{\WithFitSingle}{2907}
\FPsub{\WithFitDouble}{\WithFit}{\WithFitSingle}
\FPround{\WithFitDouble}{\WithFitDouble}{0}
\newcommand{\Qstats}{464}
\newcommand{\Pstats}{788}
\newcommand{\HIIstats}{767}
\newcommand{\PDRstats}{326}
\newcommand{\Ustats}{562}

\FPupn\AUXSUMA{\Qstats{} \Pstats{} \HIIstats{} \PDRstats{} \Ustats{} + + + +}
\FPround{\AUXSUMA}{\AUXSUMA}{0}

\ifnum\AUXSUMA=\WithFitSingle   %
\else
\AtEndDocument{\typeout{Warning: Not consistent numbers. (AUXSUMA=\AUXSUMA)}}
\fi

\FPupn\AUXSUMA{\WithFitQuiescents{} \WithFitProtostellars{} \WithFitHiiregions{} \WithFitPDRs{} \WithFitUncertain{} + + + +}
\FPround{\AUXSUMA}{\AUXSUMA}{0}

\ifnum\AUXSUMA=\WithFit %
\else
\AtEndDocument{\typeout{Warning: Not consistent numbers. (AUXSUMA=\AUXSUMA)}}
\fi

\shortauthors{Guzm{\'a}n et al.}
\shorttitle{MALT90: dust temperatures and column densities}
\newcommand{\hii}{H{\rmfamily\scshape{ii}}}

\newcommand{\Msun}{\mbox{\,$M_{\odot}$}}

\def\jnl@style{\rm}
\def\aaref@jnl#1{{\jnl@style#1}}

\def\aaref@jnl#1{{\jnl@style#1}}

\def\aj{\aaref@jnl{AJ}}                   
\def\araa{\aaref@jnl{ARA\&A}}             
\def\apj{\aaref@jnl{ApJ}}                 
\def\apjl{\aaref@jnl{ApJ}}                
\def\apjs{\aaref@jnl{ApJS}}               
\def\ao{\aaref@jnl{Appl.~Opt.}}           
\def\apss{\aaref@jnl{Ap\&SS}}             
\def\aap{\aaref@jnl{A\&A}}                
\def\aapr{\aaref@jnl{A\&A~Rev.}}          
\def\aaps{\aaref@jnl{A\&AS}}              
\def\azh{\aaref@jnl{AZh}}                 
\def\baas{\aaref@jnl{BAAS}}               
\def\jrasc{\aaref@jnl{JRASC}}             
\def\memras{\aaref@jnl{MmRAS}}            
\def\mnras{\aaref@jnl{MNRAS}}             
\def\pra{\aaref@jnl{Phys.~Rev.~A}}        
\def\prb{\aaref@jnl{Phys.~Rev.~B}}        
\def\prc{\aaref@jnl{Phys.~Rev.~C}}        
\def\prd{\aaref@jnl{Phys.~Rev.~D}}        
\def\pre{\aaref@jnl{Phys.~Rev.~E}}        
\def\prl{\aaref@jnl{Phys.~Rev.~Lett.}}    
\def\pasp{\aaref@jnl{PASP}}               
\def\pasj{\aaref@jnl{PASJ}}               
\def\qjras{\aaref@jnl{QJRAS}}             
\def\skytel{\aaref@jnl{S\&T}}             
\def\solphys{\aaref@jnl{Sol.~Phys.}}      
\def\sovast{\aaref@jnl{Soviet~Ast.}}      
\def\ssr{\aaref@jnl{Space~Sci.~Rev.}}     
\def\zap{\aaref@jnl{ZAp}}                 
\def\nat{\aaref@jnl{Nature}}              
\def\iaucirc{\aaref@jnl{IAU~Circ.}}       
\def\aplett{\aaref@jnl{Astrophys.~Lett.}} 
\def\apspr{\aaref@jnl{Astrophys.~Space~Phys.~Res.}}
\def\bain{\aaref@jnl{Bull.~Astron.~Inst.~Netherlands}} 
\def\fcp{\aaref@jnl{Fund.~Cosmic~Phys.}}  
\def\gca{\aaref@jnl{Geochim.~Cosmochim.~Acta}}   
\def\grl{\aaref@jnl{Geophys.~Res.~Lett.}} 
\def\jcp{\aaref@jnl{J.~Chem.~Phys.}}      
\def\jgr{\aaref@jnl{J.~Geophys.~Res.}}    
\def\jqsrt{\aaref@jnl{J.~Quant.~Spec.~Radiat.~Transf.}}
\def\memsai{\aaref@jnl{Mem.~Soc.~Astron.~Italiana}}
\def\nphysa{\aaref@jnl{Nucl.~Phys.~A}}   
\def\pasa{\aaref@jnl{PASA}}               
\def\physrep{\aaref@jnl{Phys.~Rep.}}   
\def\physscr{\aaref@jnl{Phys.~Scr}}   
\def\planss{\aaref@jnl{Planet.~Space~Sci.}}   
\def\procspie{\aaref@jnl{Proc.~SPIE}}   
\def\rmxaa{\aaref@jnl{Rev.~Mex. Astronom{\'i}a \& Astrof{\'i}sica}} 

\usepackage{amssymb,amsmath,natbib,graphicx}
\usepackage[T1]{fontenc}
\slugcomment{\today}

\begin{document}


\title{Far-Infrared Dust Temperatures and Column Densities of the MALT90
  Molecular Clump Sample}

\author{Andr\'es E. Guzm{\'{a}}n\altaffilmark{1,2}, Patricio
  Sanhueza\altaffilmark{3}, Yanett Contreras\altaffilmark{4,5}, Howard
  A. Smith\altaffilmark{1}, James M. Jackson\altaffilmark{6},  Sadia
  Hoq\altaffilmark{6}, and Jill M. Rathborne\altaffilmark{4}}


\altaffiltext{1}{Harvard-Smithsonian Center for Astrophysics, 60 Garden Street, Cambridge, MA, USA}
\altaffiltext{2}{Departamento de Astronom\'{\i}a, Universidad de Chile, Camino el Observatorio 1515, Las Condes, Santiago, Chile}%
\altaffiltext{3}{National Astronomical Observatory of Japan, 2-21-1 Osawa, Mitaka, Tokyo 181-8588, Japan}
\altaffiltext{4}{CSIRO Astronomy and Space Science, P.O. Box 76, Epping 1710 NSW,  Australia}
\altaffiltext{5}{Leiden Observatory, Leiden University, PO Box 9513, NL-2300 RA Leiden, the Netherlands}
\altaffiltext{6}{Institute for Astrophysical Research, Boston University, Boston, MA, USA}%
\begin{abstract}
We present dust column densities and dust temperatures for $\sim3000$ young
high-mass molecular clumps from the Millimetre Astronomy Legacy Team 90 GHz
(MALT90) survey, derived from adjusting single temperature dust emission
models to the far-infrared intensity maps measured between 160 and 870
\micron\ from the Herschel/Hi-Gal and APEX/ATLASGAL surveys.  We discuss
the methodology employed in analyzing the data, calculating physical
parameters, and estimating their uncertainties.  The population average
dust temperature of the clumps are: $16.8\pm0.2$ K for the clumps that do
not exhibit mid-infrared signatures of star formation (Quiescent clumps),
$18.6\pm0.2$ K for the clumps that display mid-infrared signatures of
ongoing star formation but have not yet developed an \hii\ region
(Protostellar clumps), and $23.7\pm0.2$ and $28.1\pm0.3$ K for clumps
associated with \hii\ and photo-dissociation regions, respectively.  These
four groups exhibit large overlaps in their temperature distributions, with
dispersions ranging between 4 and 6 K.  The median of the peak column
densities of the Protostellar clump population is $0.20\pm0.02$ gr
cm$^{-2}$, which is about 50\% higher compared to the median of the
peak column densities associated with clumps in the other
evolutionary stages.  We compare the dust temperatures and column densities
measured toward the center of the clumps with the mean values of each
clump. We find that in the Quiescent clumps the dust temperature increases
toward the outer regions and that they are associated with the shallowest
column density profiles.  In contrast, molecular clumps in the Protostellar
or \hii\ region phase have dust temperature gradients more consistent with
internal heating and are associated with steeper column density profiles
compared with the Quiescent clumps.
\end{abstract}
\keywords{stars --- stars: formation --- stars: massive, surveys -- ISM: clouds}
\vfill\eject

\section{INTRODUCTION}

Most of the star formation in the Galaxy occurs in clusters associated
with at least one high-mass star \citep{Adams2010ARA&A}.  An
understanding of star formation on global galactic and extra-galactic
scales therefore entails the study of the early evolution of high-mass
stars and how they impact their molecular environment.


The physical characterization of the places where high-mass stars form is 
an important observational achievement of the
far-infrared (far-IR) and submillimeter astronomy of the last decades. 
High-mass stars form in massive molecular clumps of sizes
$\lesssim 1$ pc, column densities $\gtrsim0.1$ gr~cm$^{-2}$, densities
$n_{\rm H_2}\gtrsim10^4$ cm$^{-3}$, and masses $>200$ \Msun\ \citep{Tan2014prpl}, with 
 temperatures depending on their evolutionary stage.  Determining the evolutionary
sequence of these massive molecular clumps and their properties is currently an active field of
study. We can define a schematic timeline that comprises four major
observational stages \citep{Jackson2013PASA,Chambers2009ApJ}: 
\begin{enumerate}
\item{Quiescent and  prestellar sources, that is, molecular clumps in the earliest
  phase with no embedded high-mass young stellar objects (HMYSOs). Some of
  these clumps are called infrared dark clumps (IRDCs) because they
  appear in absorption against the bright mid-IR background associated with
  the Galactic plane.}
\item{Protostellar clumps are those associated with signs of star
  formation such as outflows and HMYSOs, but where
  \hii\ regions have not developed.  We expect the embedded young
  high-mass stars to accrete at a high rate \citep[$\ge10^{-4}$
    \Msun\ yr$^{-1}$, e.g.,][]{McKee2003ApJ,Keto2006ApJ,Tan2014prpl}
  and to reach the main sequence in typically $\lesssim10^5$ yr
  \citep{Behrend2001AA,Molinari2008AA}.  Based on the Kelvin-Helmholtz
  contraction timescale, high-mass young stars will be likely on the
  main sequence while still accreting.}
\item{Molecular clumps associated with compact \hii\ regions. The young
  high-mass stars in these clumps have probably finished their main
  accretion phase and  have reached their final masses. 
Strong UV radiation from the newly born high-mass stars start to ionize the surrounding cocoon.}
\item{Clumps in a late evolutionary stage, where the ionizing radiation,
  winds and outflows feedback, and the expansion of the ionized gas finally
  disrupt the molecular envelope, marking the transition to an
  observational stage characterized by an extended classical \hii\ region
  and a photodissociation region (PDR).}
\end{enumerate}
Studying the dust continuum emission in the mid-IR, far-IR, and
submillimeter range is one of the most reliable ways to determine the
evolutionary phase of molecular clumps. Dust emission in the submillimeter  is usually
optically thin and traces both cold and warm environments.  By
combining large infrared Galactic plane surveys like Hi-GAL
\citep[Herschel Infrared Galactic plane survey,][]{Molinari2010PASP},
ATLASGAL \citep[APEX Telescope Large Area Survey of the
  Galaxy,][]{Schuller2009AA}, GLIMPSE \citep[Galactic Legacy Infrared
  Midplane Survey Extraordinaire,][]{Benjamin2003PASP}, and MIPSGAL
\citep{Carey2008AAS}, we can determine the evolutionary state and
calculate basic physical parameters of a large
sample of molecular clumps.

With this prospect in mind, the Millimeter Astronomy Legacy Team 90 GHz
(MALT90) survey\footnote{{Survey website: http://malt90.bu.edu/. The
    molecular line data can be accessed from
    http://atoa.atnf.csiro.au/MALT90.}} (Rathborne et al.\ in preparation;
\citealp{Jackson2013PASA,Foster2011ApJS,Foster2013PASA}) has studied
\Positions\ molecular clumps identified using SExtractor
\citep{Bertin1996AA} from the ATLASGAL data at
870\um\ \citep{Contreras2013AA,Urquhart2014AA}.  MALT90 has mapped these
clumps in 15 molecular and one  hydrogen recombination line located
in the 90 GHz atmospheric band using the 22 m Mopra telescope.  The
objective is to determine the main physical and chemical characteristics of
a statistically relevant sample of high-mass molecular clumps over a wide
range of evolutionary stages.  Approximately 80\% of the MALT90 sources
exhibit mid-IR characteristics that allow us to classify them into one of
the four preceding evolutionary stages: Quiescent, Protostellar,
\hii\ region, or PDR.  This classification of the sources was done 
  by visual  inspection of Spitzer images at 3.6, 4.5, 8.0, and 24
  \micron, as described in \citet{Hoq2013ApJ} \citep[see also][]{Foster2011ApJS}.  By combining the
MALT90 dataset with far-IR continuum and molecular line data, we can
characterize quantitatively  their temperatures, column densities, volume
densities, distances, masses, physical sizes, kinematics, luminosity, and
chemistry of the clumps.

In this paper, we focus on the dust continuum emission of the MALT90
molecular clump sample.  We model the far-IR and submillimeter emission to
derive physical parameters which, to a first approximation, are distance
independent such as the dust temperature and the column
density. Forthcoming publications by Whitaker et al. (in preparation) and
Contreras et al.\ (in preparation) will present kinematic distances and
analyze the clumps' masses, sizes, volume densities, and luminosities.
Preliminary analysis of the molecular emission indicates that the
  relative abundances, line opacities (Rathborne et al., in preparation,
  see also \citealp{Hoq2013ApJ}), and infall signatures (Jackson et al., in
  preparation) are consistent with the mid-IR classification acting as a
  proxy for clump evolution.  The MALT90  data have been already
 used in several other studies of high-mass star formation, either
based on a small ($<10$) set of relevant sources
\citep{Rathborne2014ApJ,Stephens2015ApJ,Walker2015MNRAS,Deharveng2015AA}
or using a statistical approach on a larger sample \citep[$>30$,][]{Hoq2013ApJ,Miettinen2014AA,Yu2015MNRAS,He2015MNRAS}.  In  these
studies with large samples \citep[with the exception of][]{Hoq2013ApJ}, the
dust temperature and column density of the clumps have not been
simultaneously derived from a model of the far-infrared spectral energy
distribution (SED). This paper aims to complement future high-mass star
formation studies based on the MALT90 sample by supplying robust
measurements of these physical properties and their uncertainties.


Section \ref{sec-obs} of this work presents the main
characteristics of the data set and its reduction. Section \ref{sec-ana}
describes the methods used for analyzing the data, the modeling of the dust
emission, and uncertainty and degeneracy estimations.  Section
\ref{sec-dis} discusses possible interpretations of the statistical results
of the dust parameters and, specially, how the clump evolutionary stages
correlate with the dust derived physical parameters. Section \ref{sec-sum}
summarizes the main results of this work.

{\section{OBSERVATIONS}\label{sec-obs}} 

The analysis presented in this
work is based on data taken with the \emph{Herschel Space Observatory}
\citep[HSO,][]{Pilbratt2010AA} and with the APEX telescope
\citep{Gusten2006AA}.

{\subsection{Processing of Public HSO Hi-GAL Data}\label{sec-higal}} 

We use public HSO data from the 
Herschel Infrared Galactic Plane Survey key-project
\citep[Hi-GAL,][]{Molinari2010PASP} observed between
January of 2010 and November of 2012 and obtained from the Herschel
Science Archive. The observations were made using the parallel,
fast-scanning mode, in which  five wavebands were observed
simultaneously using  the PACS \citep{Poglitsch2010AA} and the SPIRE
\citep{Griffin2010AA} bolometer arrays.  The data version
obtained from the Herschel
Science Archive corresponds to the Standard Product Generation
version 9.2.0.

Columns 1 to 4 of Table \ref{tab-ins} {list} the instrument, the
  representative wavelength in microns of each observed band, the angular
resolution represented by the FWHM of the point spread function
\citep{Olmi2013AA}, and the estimated point source sensitivity
($\sigma_p$), respectively.  The point source sensitivity, assuming
  Gaussian beams, is given by $\sigma_{\rm
    rms}\Omega_b\left(\Omega_b/2\Omega_{\rm pix}\right)^{-1/2}$, where
  $\sigma_{\rm rms}$ is the rms variations in intensity units, $\Omega_b$
  is the beam solid angle, and $\Omega_{\rm pix}$ is the pixel solid
  angle.\footnote{Theoretical justification and more detailed calculations for this formula can be found at the Green Bank Telescope technical notes:
    http://www.gb.nrao.edu/$\mathtt{\sim}$bmason/pubs/m2mapspeed.pdf
    (B. Mason, private communication)}  The fifth column gives the noise
level of the convolved and re-gridded maps (see Sections \ref{sec-noi} and
\ref{sec-conv}) and the sixth column lists the observatory where the data
were taken.  Throughout this work, we will refer to the data related to a
specific waveband by their representative wavelength in micrometers. The
position uncertainty of the Hi-GAL maps is $\sim$3\arcsec.

The generation of maps that combine the two orthogonal scan directions
was done using the Herschel Interactive Processing Environment
(HIPE) versions 9.2 and 10.  Cross-scan combination and destriping
were performed over 42 Hi-GAL fields of approximately
$2\fdg2\times2\fdg2$ using the standard tools available in HIPE. Columns 1 to 4  of Table \ref{tab-ids} give the target name, the ID of the observation, the observing mode, and the observation dates, respectively.  
For the
SPIRE maps, we applied the extended source calibration
procedure (Section 5.2.5 from the SPIRE Handbook\footnote{http://herschel.esac.esa.int/Docs/SPIRE/spire\_handbook.pdf}) 
since most of MALT90 sources correspond to dense clumps that
are comparable to or larger than the largest SPIRE beam size.
The saturation limit of
the nominal mode for SPIRE (Section 4.1.1 from the SPIRE Handbook) 
is approximately 200 Jy~beam$^{-1}$.
To prevent saturation, fields with longitudes $|l|\le5$\arcdeg\ were observed with SPIRE
using the bright observing mode instead of the nominal observing
mode.

{\subsection{Other HSO Data}\label{sec-hobys}} 

In addition to Hi-GAL data, we {used} data from three observations
made using the SPIRE bright mode by the HOBYS   key project \citep[Herschel Imaging Survey of OB YSOs,][]{Motte2010AA}.  Table
\ref{tab-ids} lists these observations' IDs. They were
directed toward the NGC 6334 ridge and the central part of M17, areas which
are heavily saturated in the Hi-GAL data.

{\subsection{ATLASGAL Archival Data}\label{sec-laboca}} 

Data at 870 \um\ were taken between 2007 and 2010 using the bolometer
LABOCA \citep{Siringo2009AA} installed on the APEX telescope located
in Chajnantor valley, Chile, as part of the ATLASGAL key project
\citep{Schuller2009AA}. Calibrated and reduced fits images were
obtained from the data public releases made by \citet{Contreras2013AA}
and \citet{Urquhart2014AA}.  Table \ref{tab-ins} displays the angular
resolution, {the point source sensitivity calculated as in Section \ref{sec-higal} using $\sigma_{\rm rms}=60$ mJy beam$^{-1}$ and $\Omega_{b}/\Omega_{\rm pix}=11.6$ \citep{Contreras2013AA}, and the  typical noise of the convolved and re-gridded ATLASGAL maps}. In addition to this noise, we assume a 10\%
uncertainty in the absolute calibration.

{\section{ANALYSIS}\label{sec-ana}}

The following sections describe the methods used in the model fitting and
uncertainty estimations.  There are \Classified\ ATLASGAL sources observed by MALT90 classified
according to their mid-IR appearance as Quiescent, Protostellar,
\hii\ region, or PDR.  The remaining sources (\Uncertain) exhibit no
clear mid-IR features that allow us to classify them unambiguously in these
evolutionary stages. We refer to these sources as ``Uncertain.''  
The
  MALT90 catalog includes \Sources\ entries, of which \Singles\ sources are
  associated with molecular emission detected at a single V$_{\rm LSR}$. 
MALT90 also detected molecular emission arising at two V$_{\rm LSR}$
  toward \DoubleClumps\ ATLASGAL sources, which correspond to
  \Doubles\ entries in the MALT90 catalog. The continuum emission from
   these sources comes from two or more clumps located at
  different distances, complicating the interpretation.  We have calculated
  column densities and temperatures toward these blended sources, but we
  have excluded them from the discussion of Section \ref{sec-dis}.

{\subsection{Noise Estimation of the HSO Data}\label{sec-noi}}

To a first approximation, the intensity assigned to each pixel is given 
 by the average of the
bolometer readings that covers that pixel position. The spatial sampling of the maps,
on the other hand, includes $\sim$3 pixels per beamwidth.  
Observed astronomical signals  vary spatially on angular scales $\gtrsim$ 1
beamsize. Therefore, in the large fraction of the map area
 that is  away from very strong sources, we expect that the differences between adjacent pixels are
dominated by instrumental noise.
In order to estimate this noise, we use the
high-pass filter defined by \citet{Rank1999IEEP} to determine the
distribution of pixel-to-pixel variations and filter out astronomical
emission. The width of this distribution determines the typical noise
through the relation $2.36\sigma=\text{FWHM}$.  The advantage
of this method is that it gives us an extra and relatively simple 
 way to estimate the noise of the final maps.  The noise estimation is similar to that obtained from  \emph{jackknife} maps, produced by taking the
difference between maps generated by the two  halves of the bolometer array \citep[see][for an analogous procedure]{Nguyen2010AA}.

The 1-$\sigma$ point source sensitivities derived from the high-pass
filter method described above are typically 18 and 24 mJy for the two
PACS bands at 70 and 160 \micron, {and 12 mJy for the three
SPIRE bands at 250, 350, and 500 \micron}. These derived
sensitivities are in good agreement with the ones expected for the
Hi-GAL survey \citep{Molinari2010PASP} and in reasonable agreement
with the sensitivities expected for the parallel
mode,\footnote{http://herschel.esac.esa.int/Docs/PMODE/html/ch02s03.html}
with the possible exception of the 160 \um\ band where we estimate
about half of the expected noise.  The noise value derived at 250
\um\ is comparable with the noise component derived by
\citet{Martin2010AA} also from Hi-GAL data, indicating that our estimation
effectively filters most of the sky emission variations, including the
cirrus noise.  Finally, and as expected, we find that the noise in
fields observed in the SPIRE bright mode is $\sim$4 times larger
compared to that in fields observed in nominal mode.  For subsequent
analyses, we consider an additional independent calibration
uncertainty of 10\% whenever we compare data among different bands, as
for example, in the SED fitting. This
10\% represents a conservative approximation of the combined
calibration uncertainty of the SPIRE photometers (5.5\%) and the beam
solid angle (4\%, see Section 5.2.13 of the SPIRE
Handbook).

{\subsection{Convolution to a Common Resolution and Foreground/Background Filtering}\label{sec-conv}}

Multi-wavelength studies of extended astronomical objects, such as
star-forming clumps and IRDCs, often combine data taken with
different angular resolutions. Therefore, to make an adequate
comparison of the observed intensities, it is necessary to transform
the images to a common angular resolution.  We accomplish this by
convolving the images to the lowest available
resolution, given by the 500 \micron\ SPIRE instrument, using the
convolution kernels of \citet{Aniano2011PASP} in the case of HSO data.
The ATLASGAL data were convolved by a two-dimensional Gaussian with FWHM
equal to $\sqrt{35\farcs0^2- 19\farcs2^2}\approx29\farcs3$, under the
 assumption that the point spread functions of the ATLASGAL and
the 500 \micron\ data are Gaussians.  In addition, to compare the
intensity of the HSO images with that of the APEX telescope, we need
to remove from the HSO data the low spatial frequency emission that
has been filtered from the ATLASGAL images.  The ATLASGAL spatial filtering is
performed during the data reduction, and is a by-product of the
atmospheric subtraction method which  removes correlated signal between
the bolometers \citep{Siringo2009AA}.  As a consequence, any uniform
astronomical signal covering spatial scales larger than 2\farcm5 is
lost \citep{Schuller2009AA}.  

We filter the HSO data in a similar way by subtracting a background image
from each field and at each band.  We assume that this background is a
smooth additive component that arises from diffuse emission either behind
or in front of the clump.  
In addition to filtering the HSO data in order to
combine it with ATLASGAL, the background subtraction serves two more
purposes: it separates the Galactic cirrus emission from the molecular
clouds \citep[e.g.,][]{Battersby2011AA}, and it corrects for the unknown
zero level of the HSO photometric scale.  
Our background model consists of a
lower-envelope of the original data under two constrains: its value at each
pixel has to be less than in the image, within a 2-$\sigma$ tolerance, and
it has to vary by less than 10\% over 2\farcm5, which corresponds to the
ATLASGAL filter angular scale.  

We construct a background image for each Hi-GAL field following a slight
modification of the \emph{CUPID-findback}
algorithm\footnote{http://starlink.jach.hawaii.edu/starlink/findback.html}
of the \emph{Starlink} suite \citep{Berry2013ASPC}.  
The iterative algorithm used to construct the background starts with
  the original image. Then, we calculate a smoothed image by setting to
  zero (in the Fourier transform plane) the spatial frequencies
  corresponding to flux variations on angular scales $<2\farcm5$.  For each
  pixel in this smoothed image with a value larger than the corresponding
  pixel in the original image plus $2 \sigma$, the pixel value from the
  smoothed image is replaced by the one in the original image, where
  $\sigma$ is the uncertainty of the map.  The remaining pixels in the
  smoothed image are kept unchanged.  The resultant map is the first
  iteration of the algorithm. This first iteration replaces the starting
  image and the cycle repeats, generating further iterations, until the
  change between two consecutive iterations is less than 5\% in all pixels.

\begin{figure}
\centering\includegraphics[width=\textwidth]{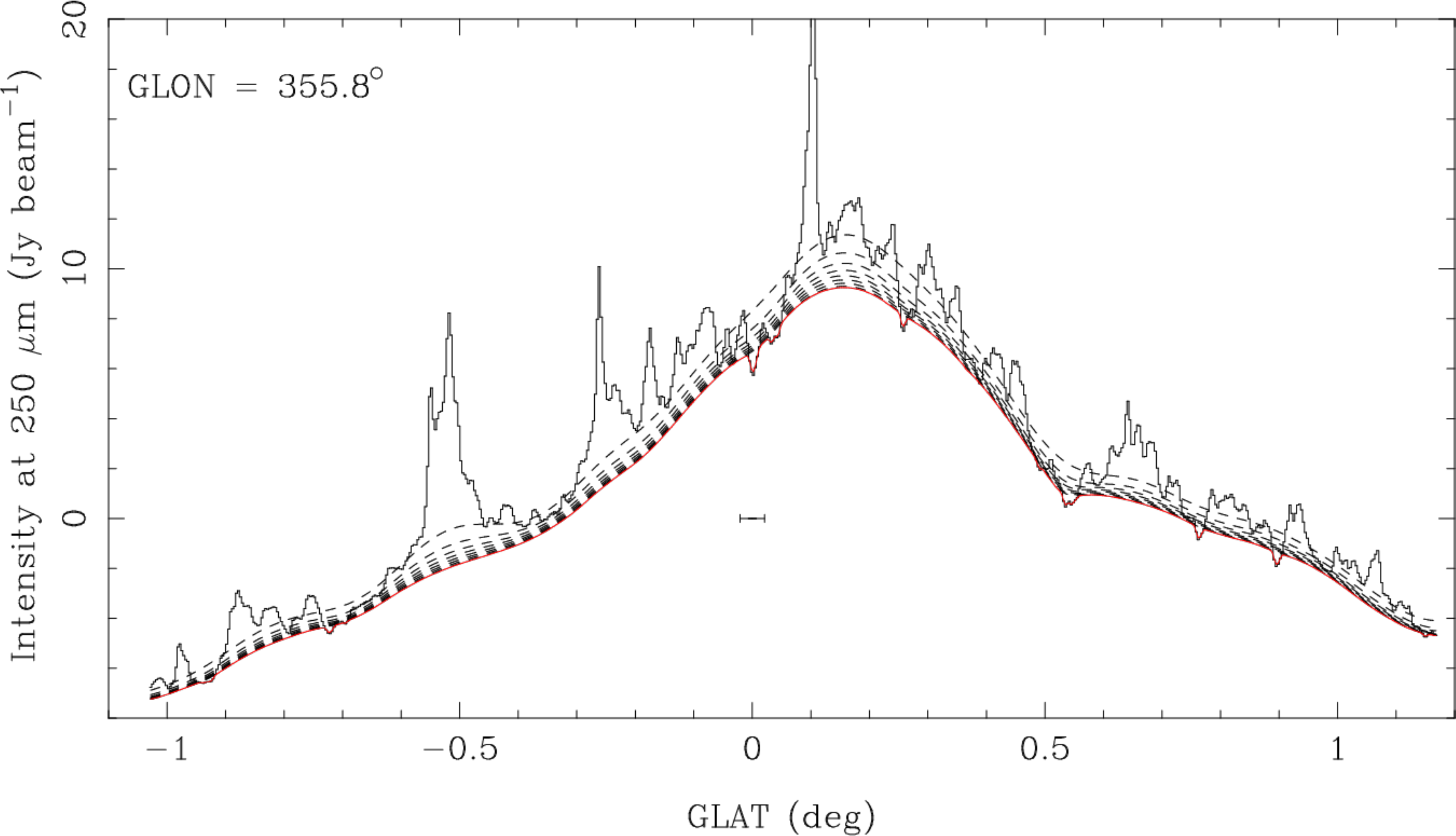}%
\figcaption{250 \um\ intensity measured across the Galactic plane
  at longitude $l=355\fdg8$. The dashed lines show the intermediate
  iterations of the background subtraction algorithm and the red line the final adopted
  background, which is a smooth lower envelope of the original data.
  The  bar located at $b=0\arcdeg$ and Intensity$=0$ Jy beam$^{-1}$ marks the shortest angular
  scale filtered by the background, that is, 2\farcm5.
Note the negative values of the intensity are due to the
  arbitrary zero-level of the HSO photometry scale.\label{fig-bac}}
\end{figure}

Figure
  \ref{fig-bac} shows an example of this process, which converges to a
  smooth lower-envelope of the original image.  The solid black line
  shows a cut along $l=355\fdg8$ of the intensity measured at 250 \um.
  Negative intensity values away from the Galactic plane are a
  consequence of the arbitrary zero-level of the HSO photometry scale.
  Dashed lines show different iterations of the algorithm and the
  final adopted background is marked in red.
The error bar at the center of
the plot measures 2\farcm5, that is, the shortest angular scale filtered by
the background.  Note that Figure \ref{fig-bac} shows a cut across latitude
at a fixed longitude, but the algorithm works on the two-dimensional image,
not assuming any particular preferred direction.

{\subsection{Single Temperature Grey-Body  Model}\label{sec-fit}}

We interpret the observed intensities as arising from a single temperature 
grey-body dust emission model. The monochromatic intensity at a frequency $\nu$ 
is given by
 \begin{equation}
I_\nu(T_d,N_g)=B_\nu(T_d)\left(1-e^{-\tau_\nu}\right)~~,\label{eq-Idust}
\end{equation}
where $B_\nu(T_d)$ is the Planck function at a dust temperature $T_d$ and 
\begin{align}
\tau_\nu&=N_{\rm dust}\kappa_\nu~~,\label{eq-tauDust}\\
        &={\rm GDR}\times N_g\kappa_\nu~~,\label{eq-gdr}
\end{align}
where $\tau_\nu$ is the dust optical depth, $N_{\rm dust}$ is the dust
column density, and $\kappa_\nu$ is the dust absorption
coefficient. The relation between the dust and gas ($N_g$)
column densities is determined by the gas-to-dust mass ratio (GDR),
which we assume is equal to $100$.  We also define the particle column
density by $N_p:=N_g/(\mu m_{\rm H})$, where
$\mu=2.3$.  The number column density of molecular hydrogen ($N_{\rm H_2}$) is
obtained in the same way but using $\mu=2.8$ \citep{Kauffmann2008AA}, under the assumption
that all the hydrogen is in molecular form. We assume throughout this work that $N_g$ is measured in gr cm$^{-2}$ and $N_{\rm H_2}$ and $N_p$ in cm$^{-2}$.
To compare the dust emission model to the data, we weight the
intensity given by Equation \eqref{eq-Idust} by the spectral response
function of the specific waveband, in order to avoid post-fitting color
corrections \citep[see for example,][]{Smith2012ApJ}.

We exclude the 70 \micron\ intensity  from the single $T_d$ fitting 
since  this emission cannot be
adequately reproduced by Equation \eqref{eq-Idust} (see Section \ref{sec-mq}).  
This problem has been noted by several authors
\citep[e.g.,][]{Elia2010AA,Smith2012ApJ,Battersby2011AA,Russeil2013AA},
who have provided at least three possible reasons:
\begin{enumerate}
\item{emission at this wavelength comes from a warmer component,}
\item{cold and dense IRDCs are seen in absorption against the Galactic plane
at 70 \um\ rather than emission,} 
\item{a large fraction of the
70 \um\ emission comes from very small grains, where the assumption
of a single equilibrium temperature is not valid.} 
\end{enumerate}
 
For each pixel and given the observed background-subtracted 
intensities $I_{\rm \nu, obs}$, 
we minimize the squared difference function,
\begin{equation}
\chi^2(T_d,N_g)=\sum_{\rm \nu}\frac{(I_{\rm \nu, obs}-\tilde{I}_{\nu})^2}{\sigma_{\nu}^2}~~,\label{eq-chi2}
\end{equation}
where the sum is taken over the observed frequencies (i.e., 5 bands) and
$\tilde{I}_{\nu}$ is the intensity spectrum predicted by the model weighted
by the respective bandpass.  The best-fit dust temperature, $T_d$, and gas
column density, $N_g$, minimize the $\chi^2$ value.  The variance
$\sigma_{\nu}^2$ is equal to the sum in quadrature of the noise (taken from
Table \ref{tab-ins}) plus 10\% of the background-subtracted intensity.  We
fit the model described in Equation \eqref{eq-Idust} for all the pixels
with intensities larger than $2\sigma_\nu$ in all bands.


The reduced $\chi^2$, defined as $\chi^2_r:=\chi_{\rm min}^2/(m-p)$
\citep{Bevington2003DRDP}, is a simple measure of the quality of the
model. Here, $\chi^2_{\rm min}$ is the minimized $\chi^2$ of Equation
\eqref{eq-chi2}, $m$ is the number of data-points, and $p$ is the number of
fitted parameters. In our case, we fit the dust temperature and the
logarithm of the gas column density, so $p=2$.  Under the hypothesis that
the data are affected by ideal, normally distributed noise, $\chi_r^2$ has
a mean value of 1 and a variance of $2/(m-p)$.

\begin{figure}
\centering\includegraphics[width=.5\textwidth]{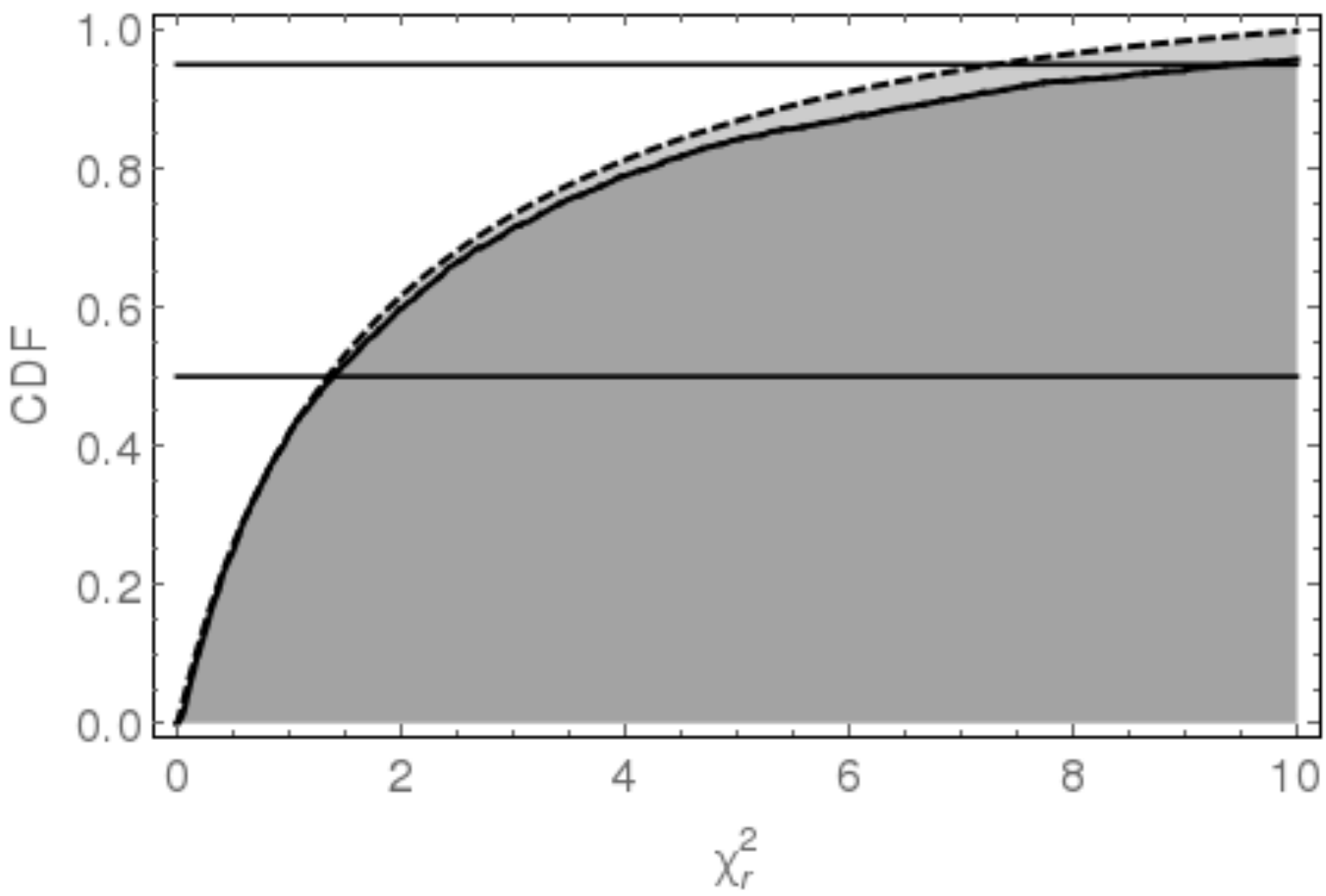}%
\figcaption{CDFs of the $\chi^2_r$.  The continuum and dashed curves
  indicate, respectively, the $\chi^2_r$ CDFs before and after the
  removal of a discordant data point (see Section
  \ref{sec-mq}). The intersection of the horizontal lines with the
  CDFs mark 0.5 (median) and 0.95 quantiles. \label{fig-chi2CDF}}
\end{figure}

Figure \ref{fig-chi2CDF} shows the $\chi^2_r$ cumulative distribution
function (CDF), calculated using all the pixels for which we fit the SED.
The median $\chi^2_r$ value is 1.6. This value is less than
$1+\sqrt{2/3}\approx1.8$, which is the expected value plus 1-$\sigma$ under
the assumption of normal errors for any particular fit.
We conclude that the SED model is in most cases adequate, or equivalently,
the limited amount of photometric data does not justify a more complicated
model.  Note that, although the distribution of $\chi^2_r$ has a reasonable
mean and median, it has a large tail: the 95\% quantile is located at
$\chi^2_r\approx9.6$.  This value represents a poor fit to the model, which
can be usually attributed to a single discordant data-point. Generally,
this point corresponds to the 870 \micron\ intensity, which illustrates the
difficulties of trying to match the spatial filtering of the HSO with
ATLASGAL data, despite the background correction and common resolution
convolution.  We re-examine the fitting when the $\chi^2_r$ value is larger
than 10 and remove from the fitting at most one data point only if its
removal decreases the $\chi^2_r$ value by a factor of 10 or more.


{\subsubsection{Spectral Index of the Dust Absorption Coefficient}\label{sec-beta}}

At frequencies $\nu<1$~THz, the dust absorption coefficient curve
$\kappa_\nu$ is well approximated by a power
law dependence on frequency  with spectral index $\beta$ \citep{Hildebrand1983QJRAS}, that is, 
\begin{equation}
\kappa_\nu=\kappa_0(\nu/\nu_0)^\beta~~.\label{eq-beta}
\end{equation}
In principle, it is possible to  quantify $\beta$ toward regions where the
emission is optically thin and the temperature is high enough such that  the
Rayleigh-Jeans (R-J) approximation is valid. In this case, from
Equations \eqref{eq-Idust} and \eqref{eq-beta} we deduce that
\begin{equation}
I_{\nu_1}/I_{\nu_2}=(\nu_1/\nu_2)^{\beta+2}~~,\label{eq-opthin}
\end{equation}
which is independent of temperature.

We estimate $\beta$ through Equation \eqref{eq-opthin} using low frequency
($<600$~GHz) data taken towards warm ($>30$ K) sources to ensure that the
R-J and the dust absorption coefficient power-law approximations are valid.
Using this value of $\beta$ we will be able to justify better the
  selection of a dust opacity law among the different theoretical
  models \citep[e.g.,][]{Ormel2011AA}.  In order to ensure that the
  sources used to estimate $\beta$ are warm enough for Equation
  \eqref{eq-opthin} to be valid, we select IRAS sources that are part of
  the 1.1 mm Bolocam Galactic Plane Survey
  \citep[BGPS,][]{Rosolowsky2010ApJS,Ginsburg2013ApJS} and the
  ATLASGAL catalog at 870 \micron. We also require that they fulfill
  $S_{60}/S_{100}>0.5$, where $S_{60}$ and $S_{100}$ are their fluxes
  at 60 and 100 \um, respectively.  
In addition,
we select sources with $|l|>10\arcdeg$ in order to avoid  possible confusion that may arise in the crowded regions around the Galactic center. 
We find 14 IRAS sources fulfilling these
requirements: 18079$-$1756, 18089$-$1837, 18114$-$1825, 18132$-$1638,
18145$-$1557, 18159$-$1550, 18162$-$1612, 18196$-$1331, 18197$-$1351,
18223$-$1243, 18228$-$1312, 18236$-$1241, 18247$-$1147, and 18248$-$1158.

The average $\beta$ calculated for these sources using Equation
\eqref{eq-opthin} is 1.6, but with a dispersion of 0.5 among the
sources. This dispersion is large, but it is compatible with a 15\%
uncertainty in the fluxes.  The spectral index is in agreement with
the absorption coefficient law of silicate-graphite grains, with
$3\times10^4$ yr of coagulation, and without ice coatings according to
the dust models from \citet{Ormel2011AA}.  For the rest of this work,
we use this model of dust for the SED fitting.  The tables compiled by
\citet{Ormel2011AA} also sample the frequency range of interest for
this work in more detail than the frequently used dust models of
\citet{Ossenkopf1994AA}.  We refrain from fitting $\beta$ together
with the SED for two reasons: i) we lack the adequate data to
effectively break the degeneracy between $\beta$ and $T_d$, that is,
good spectral sampling of highly sensitive data below $500$ GHz; 
and ii) the range of dust
models explored by fitting $\beta$ includes only power-laws instead of
using more physically motivated tabulated dust models.
 
To compare our results with previous studies, which may have
derived temperatures and column densities using different hypotheses,
we review how different assumptions on $\beta$ affect the
best-fit estimation of $T_d$. 
 Several studies
\citep[e.g.,][]{Shetty2009ApJ696-676,Shetty2009ApJ696-2234,Juvela2012AA541-33}
have discussed this problem in association with least-squares
SED fitting in the presence of noise. {They find}  that  $\beta$ and $T_d$ are
somewhat degenerate {and} associated with 
 elongated (sometimes described as banana-shaped) 
best-fit uncertainty regions in the $\beta$-$T_d$ plane. 
In this work, we stress   one aspect
that has not been sufficiently emphasized: there are \emph{two}
behaviors of the $\beta$-$T_d$ degeneracy: one is evident 
when the data cover the
SED peak, and the other when the data only cover the
 R-J part of the spectrum.  In the first case the
degeneracy is well described by the modified Wien displacement law
\begin{equation}
\frac{h\nu_{\rm peak}}{k T_d}\approx(\beta+3)~~,\label{eq-mwdl}
\end{equation}
that is, the uncertainty region of $T_d$ and $\beta$ is elongated along the
curve defined by Equation \eqref{eq-mwdl}.  In Equation \eqref{eq-mwdl},
$\nu_{\rm peak}$ represents the frequency where the SED takes its maximum value, which under optically thin
conditions is proportional to the temperature. The proportionality constant
depends on $\beta$ in a complicated way, but the approximation of Equation
\eqref{eq-mwdl} is  correct within a 10\% for $\beta>1$ and within a 20\%
for all $\beta\ge0$.  Note that by assuming a value of $\beta$ and determining
$\nu_{\rm peak}$ observationally we can estimate $T_d$ using Equation
\eqref{eq-mwdl} in a simple way.  \citet{Sadavoy2013ApJ} and \citet[][their
  20 K case]{Shetty2009ApJ696-676} show examples of uncertainty regions
given by the iso-contours of the $\chi^2$ function which are elongated
along the curve defined in Equation \eqref{eq-mwdl}.  On the other hand, if
the spectral range of the data does not cover the observed peak of the SED
and covers only the R-J region, the degeneracy between $\beta$ and $T_d$ is
better described by the following relation,
\begin{equation}
\beta-\frac{h\nu_m}{2 k T_d}= {\rm constant}~~,\label{eq-RJ}
\end{equation}
where $\nu_m$ is the highest observed frequency.  This relation describes
well the degeneracy of the high temperature curves (60 and 100 K) shown in
\citet{Shetty2009ApJ696-676}.  The constant in the right hand side of
Equation \eqref{eq-RJ} is approximately
\[2+\frac{d\ln S_{\nu_m}}{d \ln \nu}~~,\] that is, 2 plus the logarithmic 
derivative (or spectral index) of the spectrum evaluated in the
highest observed frequency.  In practice, the exact value of the
constants in the right hand side of Equations \eqref{eq-mwdl} and
\eqref{eq-RJ} can be determined from the best-fit
solutions.  In this work, the HSO bands usually cover the peak of the
SED so Equation \eqref{eq-mwdl} is more pertinent.
Depending on the spectral sampling, we can use Equation \eqref{eq-mwdl} or
\eqref{eq-RJ} to compare temperatures between studies that assume
different values of $\beta$.  
For example, the emission in the HSO
bands from a cloud of $T_d=15$~K with a $\beta=1.0$ dust absorption
law is also consistent, by Equation \ref{eq-mwdl} and assuming 10\%
uncertainty, with the emission coming from a cloud of $T_d=12$~K and
$\beta=2$. In each case, the HSO bands cover the peak
of the SED. We use this method to re-scale and compare 
best-fit temperatures obtained from the literature in Section \ref{sec-dis}.

{\subsection{Model Uncertainties}\label{sec-mq}} 

\begin{figure}
\centering\includegraphics[width=.5\textwidth]{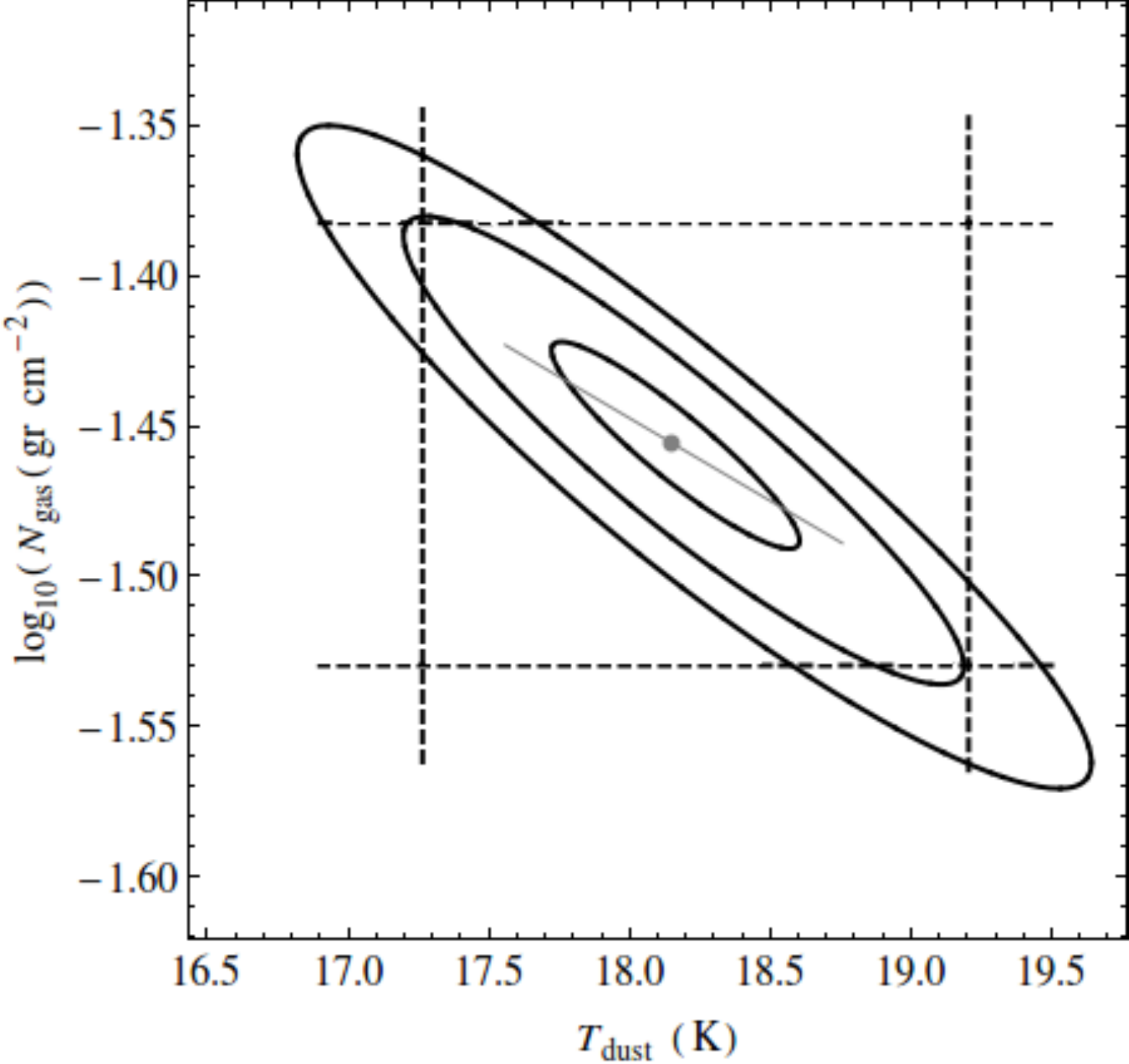}%
\figcaption{Contour map of $\Delta\chi^2=\chi^2-\chi^2_{\rm min}$ for
  an arbitrary pixel around $\Delta\chi^2=0$.  Contour levels are
  0.45, 2.3, and 4.61, corresponding to 0.25, 1.0, and 1.6 $\sigma$,
  respectively.  Dashed lines display the predictions of
  Equations \eqref{eq-unc} which approximate the 1-$\sigma$ contour
  projection onto the axes.  The thin gray line that crosses the best
  fit point has the direction of the curve $N_gT_d=\text{constant}$. \label{fig-Dchi2con}}
\end{figure}

We estimate the best-fit parameter uncertainties using the projection
of the 1-$\sigma$ contour of the function
$\Delta\chi^2:=\chi^2-\chi^2_{\rm min}$ \citep{Lampton1976ApJ}. In the
case of 2 fitted parameters, the 1-$\sigma$ uncertainty region is enclosed by the
$\Delta\chi^2=2.3$ contour.  The parameter uncertainties for the SED
fitting are given by the projections of these uncertainty regions onto
the $T_d$ and $\log N_{g}$ axes.  For pixels in images observed using
the nominal observing mode, the projections are well described by
the following equations
\begin{equation}
\label{eq-unc}%
\begin{aligned}
\delta T^{-} &=\eta_{10}\left(0.3-0.4~T_{10}+ 0.4~T_{10}^2\right)~~,\\
\delta T^{+} &= \eta_{10}\left(1.1-1.3~T_{10}+0.7~T_{10}^2\right)~~,\\
\delta\log N_g &= \eta_{10}\left(0.03-0.03\log N_g\right)~~, 
\end{aligned}
\end{equation}
where $T_{10}=T_d/(10~{\rm K})$, $N_g$ is in gr~cm$^{-2}$, and
$\eta_{10}$ is the flux calibration uncertainty in units of 10\%. 
The best fit temperature and log-column density with 
 their 1-$\sigma$ uncertainties are given by ${T_d}^{+\delta T^{+}}_{-\delta T^{-}}$ and 
$\log N_g\pm\delta \log N_g$, respectively.  
For pixels in images observed using the bright observing 
mode, the projections are well described by 
\begin{equation}
\label{eq-uncB}%
\begin{aligned}
\delta T^{-} &=\eta_{10}\left(0.7-0.71~T_{10}+ 0.53~T_{10}^2\right)~~,\\
\delta T^{+} &=\eta_{10}\left(1.1-1.3~T_{10}+0.74~T_{10}^2\right)~~,\\
\delta\log N_g &= \eta_{10}\left(0.05-0.03\log N_g\right)~~.
\end{aligned}
\end{equation}
Equations \eqref{eq-unc} and \eqref{eq-uncB} were derived by fitting
the upper and lower limits of $T_d$ and $\log N_g$ projections of the
uncertainty region.  
  These approximations for the uncertainty are
valid for $T_d$ between 7 and 40 K, for $\log N_g $ between $-3.4$ and
$1.1$ (equivalent to $\log N_{\rm H_2} $ between 19.9 and 24.4), and for
values of $\eta_{10}$ between 1 and 2, which correspond to 10\%
and 20\% calibration errors, respectively.  Figure \ref{fig-Dchi2con}
shows an example of the  prediction of Equations \eqref{eq-unc}
compared to the $\Delta\chi^2$ contours.  Within their range of
validity and for data taken in the nominal mode, the intervals 
$\left[T_d-\delta T^{-},T_d+\delta T^{+}\right]$ and 
$\left[\log N_g-\delta \log N_g,\log N_g+\delta \log N_g\right]$ 
correspond to the projections of the uncertainty ellipse onto the 
 $T_d$ and $\log N_g$ axes
within 0.2 K and 0.02 dex, respectively.

Equations \eqref{eq-unc} also indicate that, while the confidence
interval of $\log N_p $ is symmetric and roughly constant, the
temperature uncertainties grow rapidly above 25 K and they are skewed
towards the higher value \citep{Hoq2013ApJ}. This is produced   mainly because of the absence of data at wavelengths shorter than 160 \um.  
As explained in Section \ref{sec-fit}, we do not use the 70
\micron\ data in our model.  Including the 70
\um\ data increases the median of the $\chi^2_r$ distribution to $\sim3.5$.
The temperature and
log-column density uncertainties are also correlated along the approximate direction
where the product $T_d\times N_g $ is constant, indicated in Figure
\ref{fig-Dchi2con}. The better the R-J approximation is for the SED, the
better will be  the alignment of the major axis of the ellipse with the line
$T_d\times N_g=\text{constant}$.

{\subsection{Saturated Sources}\label{sec-sat}}

The HSO detectors in SPIRE and PACS have saturation limits that depend on
the observing mode.  The saturation intensities for the
nominal observing mode are 220 and 1125 Jy~beam$^{-1}$ for the 70 and
160 \um\ PACS bands\footnote{PACS Observer's Manual, v.\ 2.5.1, Section 5.4}, 
respectively, and 200 Jy~beam$^{-1}$ for SPIRE.
Saturation is most problematic in the 250 \um\ SPIRE band.  

There are 46 MALT90 sources whose Hi-GAL data are affected by
saturation.  Of these, six are covered by HOBYS observations made
by using the bright mode (Section \ref{sec-hobys}) that gives  reliable 250
\um\ intensities.  For the remaining 40 sources, we replace the
saturated pixels with the saturation limits given above, and we fit
the SED taking these values as lower
bounds.

{\subsection{The 70 $\mu$m Appearance of the Quiescent Clumps}\label{sec-re70}}

MALT90 and other previous studies
\citep[e.g.,][]{Molinari2008AA,Lopez-Sepulcre2010AA,Sanhueza2012ApJ,Sanchez-Monge2013MNRAS,Giannetti2013AA,Csengeri2014AA}
use mid-IR observations as a probe of star formation activity.  However,
deeply embedded, early star formation activity could be undetected at
mid-IR yet be conspicuous at far-IR.  Quantitatively, we expect that the 24
to 70 \um\ flux density ratio of HMYSOs to vary between $10^{-6}$ and 1 for
a wide range of molecular core masses (60 to 240 \Msun) and central star
masses over 1~\Msun\ \citep{Zhang2014ApJ}.  Therefore, despite MIPSGAL
having $\sim 50$ times better point source sensitivity at 24 \micron\ than
that of Hi-GAL at 70 \micron, it is possible to detect embedded protostars
in 70 \um\ images that would otherwise appear dark at 24 \um\ and would be
classified as Quiescent.  The 70 \um\ data thus allow us to further refine
the MALT90 classification since a truly Quiescent clump should lack 70 \um\
compact sources \citep[e.g.,][]{Sanhueza2013ApJ,Beuther2013AA553,Tan2013ApJ}.

We examined the Hi-GAL 70 \um\ images of the \WithFitQuiescents\ Quiescent sources
and found 91 (15\%) that show compact emission at 70 \um\ within
38\arcsec\ -- or one Mopra telescope beamsize -- of the nominal MALT90
source position.  Hereafter, we consider these sources as part of the Protostellar sub-sample.  We also
found 83 sources that appear in absorption at 70 \um\ against the
diffuse Galactic emission.  We refer to these clumps as far-IR dark
clumps (far-IRDCs). The remaining 442 Quiescent sources are either associated with diffuse emission not useful for tracing embedded star formation, or they are confused with the 70 \um\ diffuse emission from the Galactic plane.

{\section{DISCUSSION}\label{sec-dis}}

\begin{figure}
\includegraphics[width=0.84\textwidth]{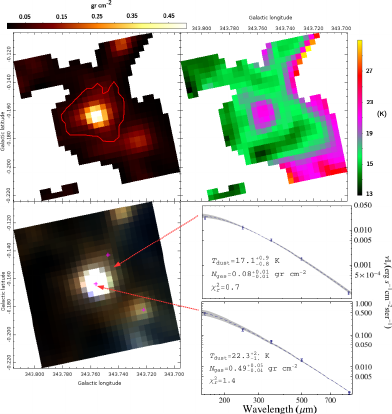}
\caption{\small Parameters and data around AGAL343.756$-$00.164. Bottom
  panels show a three color image (500, 250, and 160 \um\ for red, green,
  and blue, respectively) centered on AGAL343.756$-$00.164 and two SED
  fittings performed over the intensities measured toward two positions,
  marked by red arrows: one located toward the center of the MALT90 source
  and the other toward its periphery. The three crosses mark the position of
  adjacent MALT90 sources. The top-left panel shows the best-fit column
  density map. The red contour marks the area that defines the
  AGAL343.756$-$00.164 clump.  Blanked pixels do not have intensities
  larger than $2 \sigma$ in at least one of the bands (see Section
  \ref{sec-fit}).  The top-right panel shows the best-fit dust temperature
  map.\label{fig-sed}}
\end{figure}

Figure \ref{fig-sed} shows the dust temperature and column density obtained
for each pixel around the source AGAL343.756$-$00.164, which is taken as a
typical example. Best-fit dust temperatures, column densities, and their
uncertainties are calculated pixel by pixel. The two plots located in the
lower-right corner of Figure \ref{fig-sed} show the SED measured in two
directions (center and periphery) toward AGAL343.756$-$00.164. The blue
dashed line in each of these plots is the curve given by Equation
\eqref{eq-Idust} evaluated in the best fit solution. The shaded region
around the curve is the locus covered by the model when the best-fit
parameters vary within the 1-$\sigma$ confidence interval.  As explained in
Section \ref{sec-fit}, the $\chi^2$ is calculated by comparing the measured
intensities at each band with the SED model weighted by the respective
bandpasses.


Table \ref{tab-NT} gives the derived dust temperatures and log-column densities
of \WithFit\ MALT90 sources. 
This correspond to a 99.1\% of the \Positions\ ATLASGAL sources
  observed by MALT90. The remaining \NoFit\ sources are either not covered
  by HSO observations (24 sources) or they are too faint to reliably estimate
  the dust parameters (4 sources).
Column 1 indicates the ATLASGAL name of the
source. We include \WithFitDouble\ entries which
correspond to multiple sources blended along the same line of sight,
indicated with an ``m'' superscript.  
Column 2 {gives} the effective angular radius of the source in arcsec,
defined as $\theta_{\rm eff}=\sqrt{\Omega_s/\pi}$, where $\Omega_s$ is the effective angular
area occupied by the MALT90 source.  This area corresponds to the
intersection between the region enclosing the source where the column
density is greater than 0.01~gr~cm$^{-2}$ ($>2.0\times10^{21}$~cm$^{-2}$ in
\hh\ column density) and the 870 \um\ ATLASGAL mask (see
\citealp{Contreras2013AA} and \citealp{Urquhart2014MNRAS}). Figure
\ref{fig-sed} shows an example of one of these areas (red contour in top left image).
Column 3 of Table \ref{tab-NT} {gives} the mean dust temperature averaged over the area of each
source ($\bar{{T_d}}$).
Columns 4 and 5 {list} the lower and upper uncertainty of $\bar{T_d}$,
respectively.
Columns 6, 7, and 8 give the dust temperature at the position of the 
870 \um\ peak intensity  ($T_{d,{\rm P}}$) and its lower and upper uncertainties,
respectively.
Columns 9, 10, and 11 list the  average column density ($\bar{N_{g}}$), its logarithm, and the 
uncertainty of the latter, respectively. 
Columns 12 gives the peak column density ($N_{g,{\rm P}}$), derived using 
the 870 \um\ peak intensity and $T_{d,{\rm P}}$ (in  Equations \eqref{eq-Idust}, \eqref{eq-tauDust} and \eqref{eq-gdr}).
Columns 13 and 14 give $\log N_{g,{\rm P}}$ and its uncertainty,
respectively.
Finally, column 15 {gives} the mid-IR classification of the MALT90 source, as
Quiescent (\WithFitQuiescents\ clumps), Protostellar (\WithFitProtostellars\ clumps), \hii\ region (\WithFitHiiregions\ clumps), PDR (\WithFitPDRs\ clumps), or Uncertain (\WithFitUncertain\ clumps). Note that these numbers describe the statistics of Table \ref{tab-NT}, that is, of the \WithFit\ sources for which we have dust column density and temperature estimations.\
  For the Quiescent
sources, we indicate with a superscript ``C'' or ``D'' whether the source
is associated with 70 \um\ compact emission or if it is a far-IRDC,
respectively (see Section \ref{sec-re70}). 
No superscript means that neither of these features appears related to the clump.

Previous studies of massive molecular clumps have relied on samples
obtained from the IRAS catalog and fit SEDs to obtain dust temperatures and
masses.  We find a total of 116 matches between MALT90 and those samples
as analyzed by \citet[94 matching sources]{Faundez2004AA} and \citet[22
  matches]{Giannetti2013AA}.  
Other studies, such as
\citet{Sridharan2002ApJ} and \citet{Williams2005AA}, have targeted the
northern sky and they do not overlap significantly with MALT90 (1 source in common each).
%
From all these sources, 63 are
classified as  \hii\ region, 44 as Protostellar, 3 as PDR, 6 as
Quiescent, and 2 as Uncertain.  From the relative fraction of Quiescent sources
in the MALT90 sample, we would expect 13 or more  of the 118 to be
Quiescent with a 99\% probability, assuming that they are randomly
sampled. Since there are only 6 Quiescent matches, we conclude that
previous surveys were  biased toward more evolved stages, illustrating how MALT90 helps to fill in  the gap in the study of
cold  clumps.

\begin{figure}
\includegraphics[width=\textwidth]{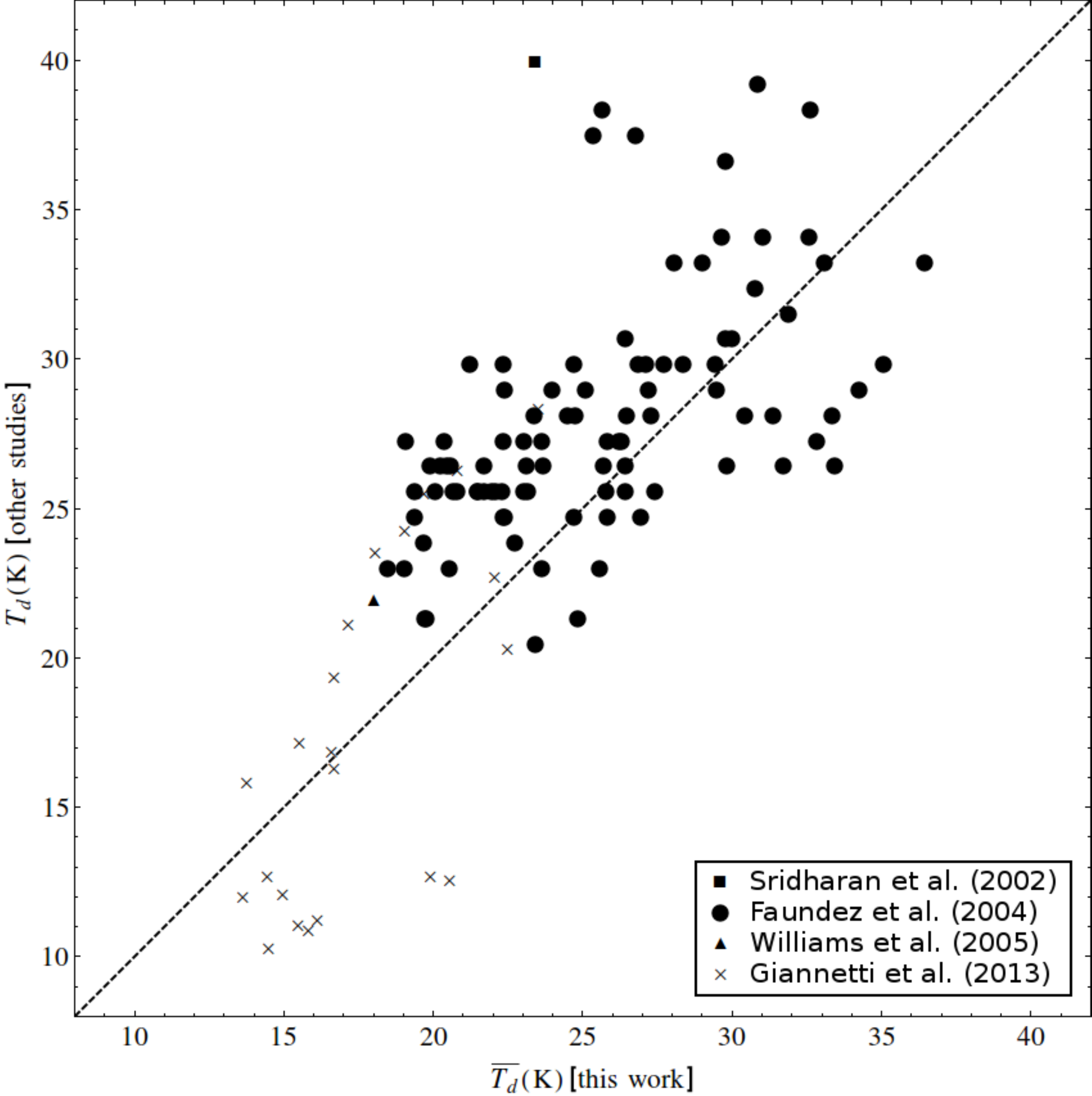}
\figcaption{Comparison between the dust temperatures obtained in the
  present work and the dust temperatures obtained from
  \citet{Sridharan2002ApJ}, \citet{Faundez2004AA},
  \citet{Williams2005AA}, and \citet{Giannetti2013AA}. The dust
  temperatures of \citet{Faundez2004AA} and \citet{Giannetti2013AA}
  have been re-scaled  according to the assumed value of $\beta$
  using Equation \eqref{eq-mwdl}. The dashed line shows the identity line.
\label{fig-comp}}
\end{figure}

Figure \ref{fig-comp} shows the dust temperature calculated by previous
studies versus the dust temperatures given in this work. We calculate a
Spearman correlation coefficient \citep[Section 4.2.3]{Wall2012psa} of 0.75
with a 95\% confidence interval between 0.68 and 0.83, indicating a
positive correlation between our temperature estimations and those from the
literature.  \citet{Faundez2004AA} assume a dust absorption spectral index
$\beta=1$, while our dust model is characterized by
$\beta\approx1.7$. Therefore, we correct their temperatures according to
Equation \eqref{eq-mwdl} by multiplying them by $(3+1)/(3+1.7)\approx0.85$.
The correction decreases the mean of the differences between the dust
temperatures obtained by \citet{Faundez2004AA} and the temperatures
obtained by us from $+7$ K (uncorrected) to $+2$ K.  We apply the same
correction to the temperatures given by \citet{Giannetti2013AA} using their
reported best-fit $\beta$ values.  Figure \ref{fig-comp} shows that
temperatures estimated using data from mid- and far-IR bands below 100
\um\ are more often higher than the dust temperatures derived in this work.
In consequence, the slope of a linear regression performed in the data shown in Figure \ref{fig-comp} is slightly larger than unity ($1.13\pm0.09$).
This
is somewhat expected since our own single temperature SED model
underestimates the 70 \um\ intensity (see Section \ref{sec-fit}).  The most
plausible reason is that in these sources there is a warmer dust component
better traced by IR data below 100 \um.


Dust temperatures and column densities of a preliminary MALT90 subsample
consisting of 323 sources were presented by
\citet{Hoq2013ApJ}.\footnote{\citet{Hoq2013ApJ} report 333 sources, but
  only 323 of these are part of the final catalog.}  They also use the
Hi-GAL data (without ATLASGAL) and they employ a similar data processing
and SED fitting procedure compared to that used in this
work. \citet{Hoq2013ApJ} report dust temperatures which are consistent
within 13\% compared to the ones given in Table \ref{tab-NT}.  However, we
obtain average column densities that are smaller by about 20\%.  The
differences are due to our source sizes being larger than the ones assumed
by \citet{Hoq2013ApJ}.  They use a fixed size equal to one Mopra telescope
beam, while we define the size of the source based on its extension in the
column density map.

When there is more than one source in the same line of sight, the continuum
emission blends two or more clumps located at different distances. This
makes uncertain the interpretation of the temperature, column density,
and evolutionary stage classification.  Therefore, for further
analysis we remove these sources from the MALT90 sample, leaving
\WithFitSingle\ sources.  This number breaks down in the following
  way (see Table \ref{tab-means}): there are \Qstats\ sources considered as Quiescent (single V$_{\rm
    LSR}$, without a compact 70 \um\ source), \Pstats\ considered
  Protostellar (including  Quiescents with a 70 \um\ compact source),
  \HIIstats\ \hii\ regions, and \PDRstats\ PDRs. The remaining sources (\Ustats) have an
  Uncertain classification.  This selection and reclassification of
sources, as we show in Appendix \ref{sec-stat}, does not affect the
conclusions presented in the following sections.

\subsection{Dust Temperature versus Gas Temperature}

\begin{figure}
\includegraphics[width=0.95\textwidth]{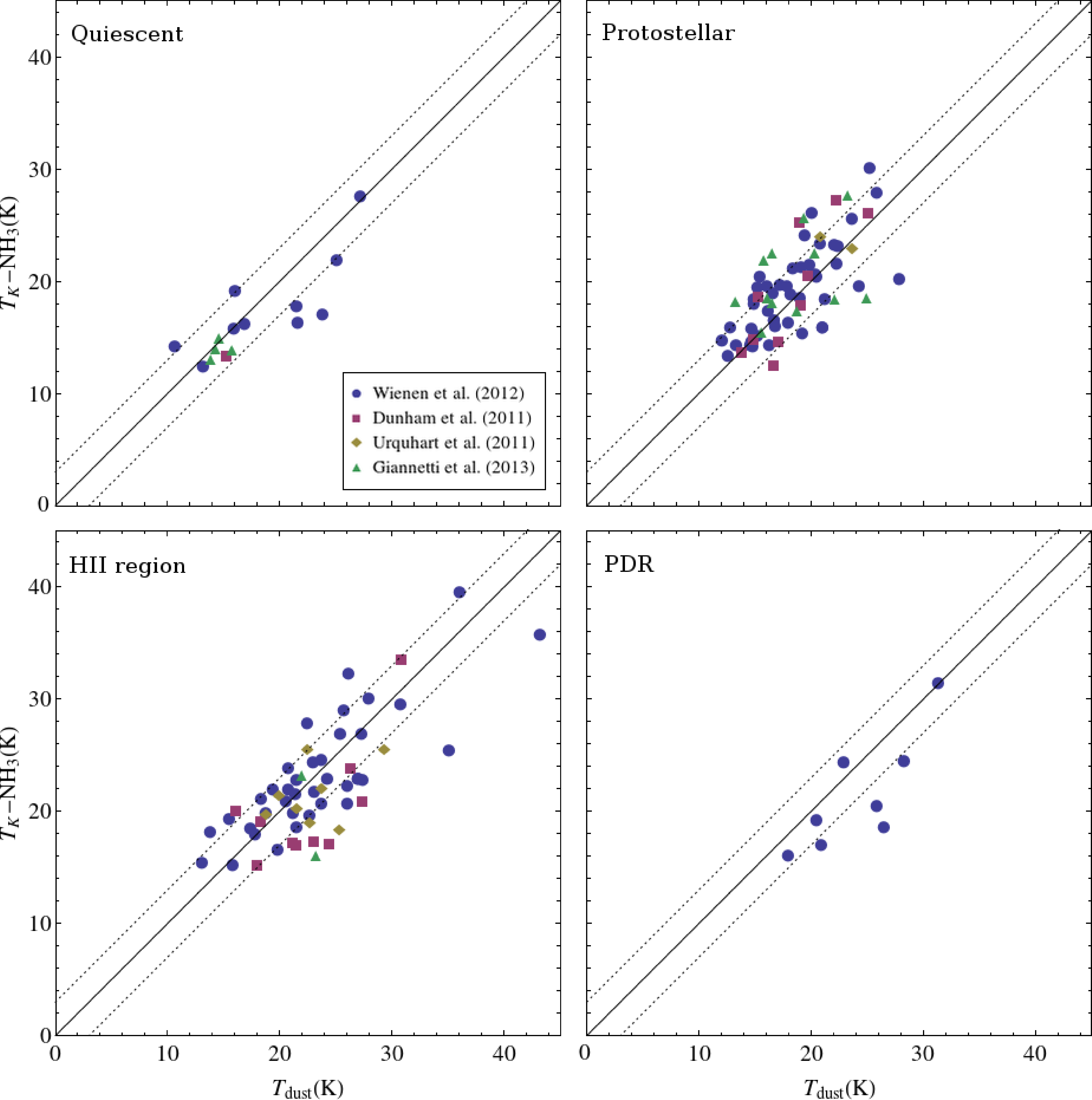}
\figcaption{{From left to right and top to bottom: dust temperature
    versus kinetic ammonia temperature of 15 Quiescent, 69 Protostellar, 58
    \hii\ region, and 8 PDR MALT90 sources.  Ammonia temperatures were
    obtained from \citet{Wienen2012AA}, \citet{Dunham2011ApJ},
    \citet{Urquhart2011MNRAS}, and \citet{Giannetti2013AA}.  The continuous
    and both dotted lines in each panel display the identity and the
    identity $\pm3$~K relations, respectively.\label{fig-amm}}}
\end{figure}

The dust temperature is fixed by the balance between heating
and radiative cooling of the grain population.  If the density in a
molecular cloud is greater than $5\times10^4$~cm$^{-3}$
\citep{Goldsmith2001ApJ,Galli2002AA}, we expect the dust temperature to be
coupled to the gas temperature.  
We test this hypothesis by comparing the
average dust temperatures with the gas kinetic temperatures determined  from
ammonia observations.  Figure \ref{fig-amm} shows the ammonia temperature
derived by 
\citet[23 matching sources]{Dunham2011ApJ}, 
\citet[{10 matching  sources}]{Urquhart2011MNRAS}, 
\citet[106 matching sources]{Wienen2012AA},
and \citet[19 sources]{Giannetti2013AA} 
versus the dust temperature, separated by evolutionary stage.
\citet{Dunham2011ApJ} and \citet{Urquhart2011MNRAS} performed the NH$_3$
observations using the Green Bank Telescope at 33\arcsec\ angular
resolution.  \citet{Wienen2012AA} used the Effelsberg Radiotelescope at
40\arcsec\ angular resolution.  These are comparable to the resolution of
our dust temperature maps (35\arcsec).  On the other hand, \citet{Giannetti2013AA} used
NH$_3$ data obtained from ATCA with an angular resolution of
$\sim20\arcsec$. In this last case, we compare their ammonia temperatures
with $T_{d,{\rm P}}$ instead of $\bar{T_d}$.

All but eight MALT90 sources with ammonia temperature estimation are
classified in one of the four evolutionary stages. The Spearman correlation
coefficient of the entire sample (158 sources, including these eight with
Uncertain mid-IR classification) {is 0.7, with a 95\% confidence interval
between 0.6 and 0.8,} indicating a positive correlation between both
temperature estimators.
The scatter
of the relation is larger than the typical temperature uncertainty, and it
grows with the temperature of the source.  For sources below 22 K, ammonia
and dust temperatures agree within $\pm3$ K. Above 22 K, the uncertainties
of both temperature estimators become larger \citep[see Equations
  \eqref{eq-unc} and, for example,][]{Walmsley1983AA}, consistent with the
observed {increase of the scattering.}
In addition, higher temperature clumps are
likely being heated from inside and therefore associated with more
variations in the dust temperatures along the line of sight, making the
single temperature approximation less reliable.  
{The slopes of the linear regressions performed in the data are
  $0.7\pm0.1$, $0.8\pm0.1$, $0.7\pm0.1$, and $0.9\pm0.3$ for the Quiescent,
  Protostellar, \hii\ region, and PDR samples, respectively.}
%
The 
ammonia and dust temperatures relation agrees in general  with that found by
\citet{Battersby2014ApJ786}, except that we do not find a systematically
worse agreement on Quiescent sources compared with the other evolutionary
stages.

\subsection{Temperature and Column Density Statistics}

\begin{figure}
\includegraphics[width=.87\textwidth]{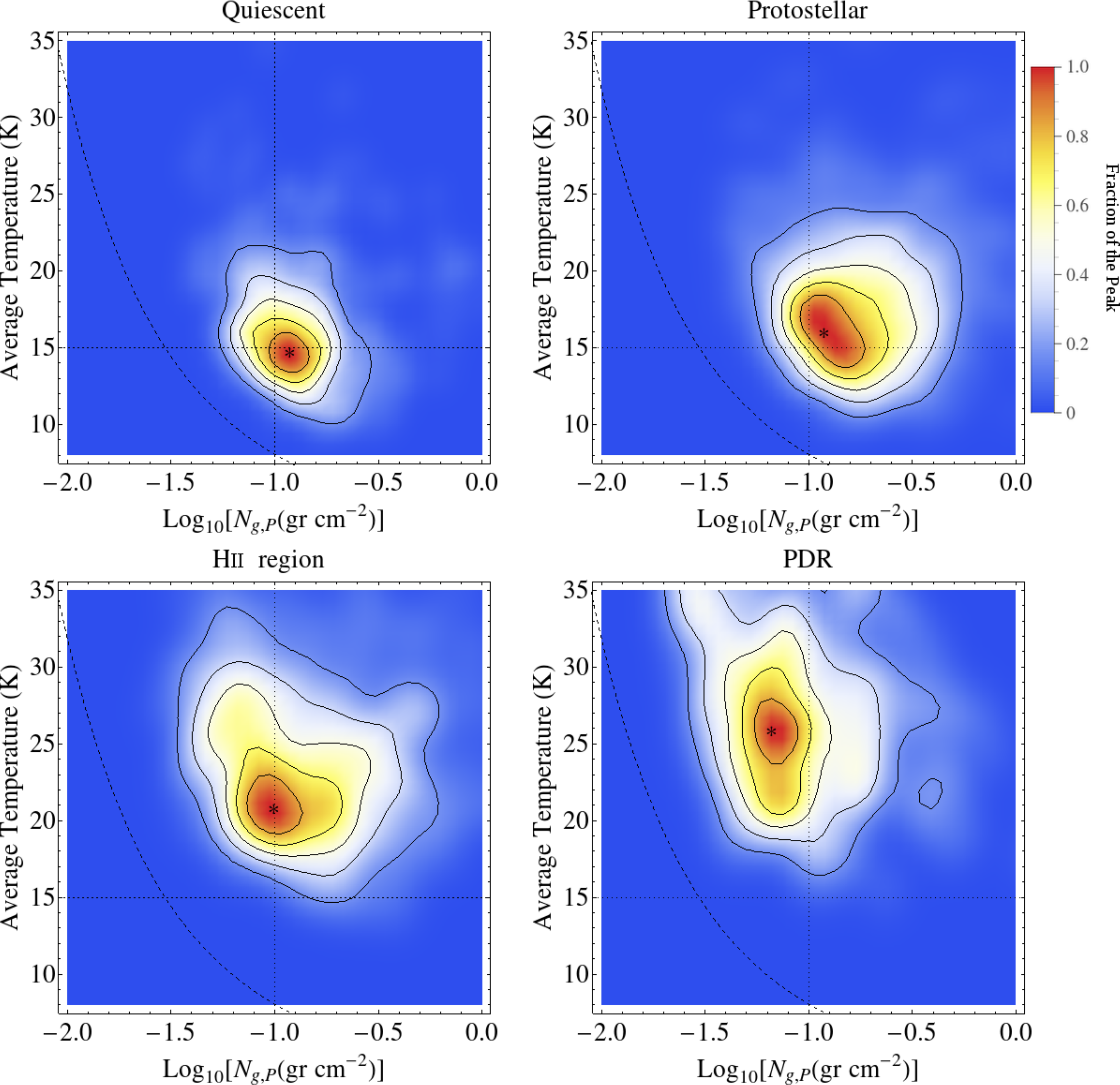}
\figcaption{\small Color map of smoothed 2-D histograms of the distributions of
  temperature and log peak column density for each evolutionary
  stage.  The smoothing was implemented using a Gaussian kernel with a
  standard deviation of 25\% of the observed dispersion in each
  direction. The color scale  indicates the fraction of the peak according to the colorbar. Five contours are taken from one to five sixths of the peak. 
  An asterisk shows the position of the peak of each distribution.  The
  dashed line indicates the 0.5 Jy level at 870 \um\ (assuming the dust model described in Section \ref{sec-fit}), which is
  roughly the completeness limit of MALT90 \citep{Jackson2013PASA}. The two
  dotted lines drawn at $\log(N_{g,{\rm P}})=-1.0$ and $\bar{T_d}=15$ K are 
  fiducial values used to facilitate the comparison among the different panels.
\label{fig-sd}}
\end{figure}

Figure \ref{fig-sd} shows {maps} of smoothed 2-D histograms of
the distributions of $\bar{T_d}$ and $\log N_{g,{\rm P}}$ of the
MALT90 clumps for each mid-IR classification.  
In the following analysis we focus on these two quantities and their
  relation with the evolutionary stage. We use $N_{g,{\rm P}}$ instead of
  $\bar{N_g}$ because $N_{g,{\rm P}}$ is independent of the specific
  criterion used to define the extension of the clump and because the
  column density profiles are often steep \citep[$\propto s^{-0.8}$, where
    $s$ is the plane-of-the-sky distance to the clump
    center,][]{Garay2007ApJ}, making the average $\bar{N_g}$ less
  representative of the clump column density values.  On the other hand,
  dust temperature gradients are shallower \citep[$\propto r^{-0.4}$, where
    $r$ is the distance to the clump center, {see}][]{vanderTak2000ApJ} and
  $\bar{T_d}$ has less uncertainty compared with the temperature calculated
  toward a single point.
%
We include in the
Protostellar group those sources that are associated with a
compact source at 70 \um\ (Section \ref{sec-re70}).
The most conspicuous differences between the evolutionary stages 
are evident between  the Quiescent/Protostellar 
and the \hii\ region/PDR populations
\citep{Hoq2013ApJ}.  The  main difference between these groups is the
temperature distribution.  Most of the sources in the
Quiescent/Protostellar stage have temperatures below 19 K, while most
\hii\ region/PDR sources have temperatures above 19 K.  We also note that 
the Quiescent, Protostellar, and \hii\ region populations have peak column densities  $\gtrsim0.1$~gr~cm$^{-2}$, equivalent to
$2.13\times10^{22}$ \hh\ molecules per cm$^{2}$, while the PDR population has peak column densities of typically half of this value.

\begin{figure}
\includegraphics[width=1\textwidth]{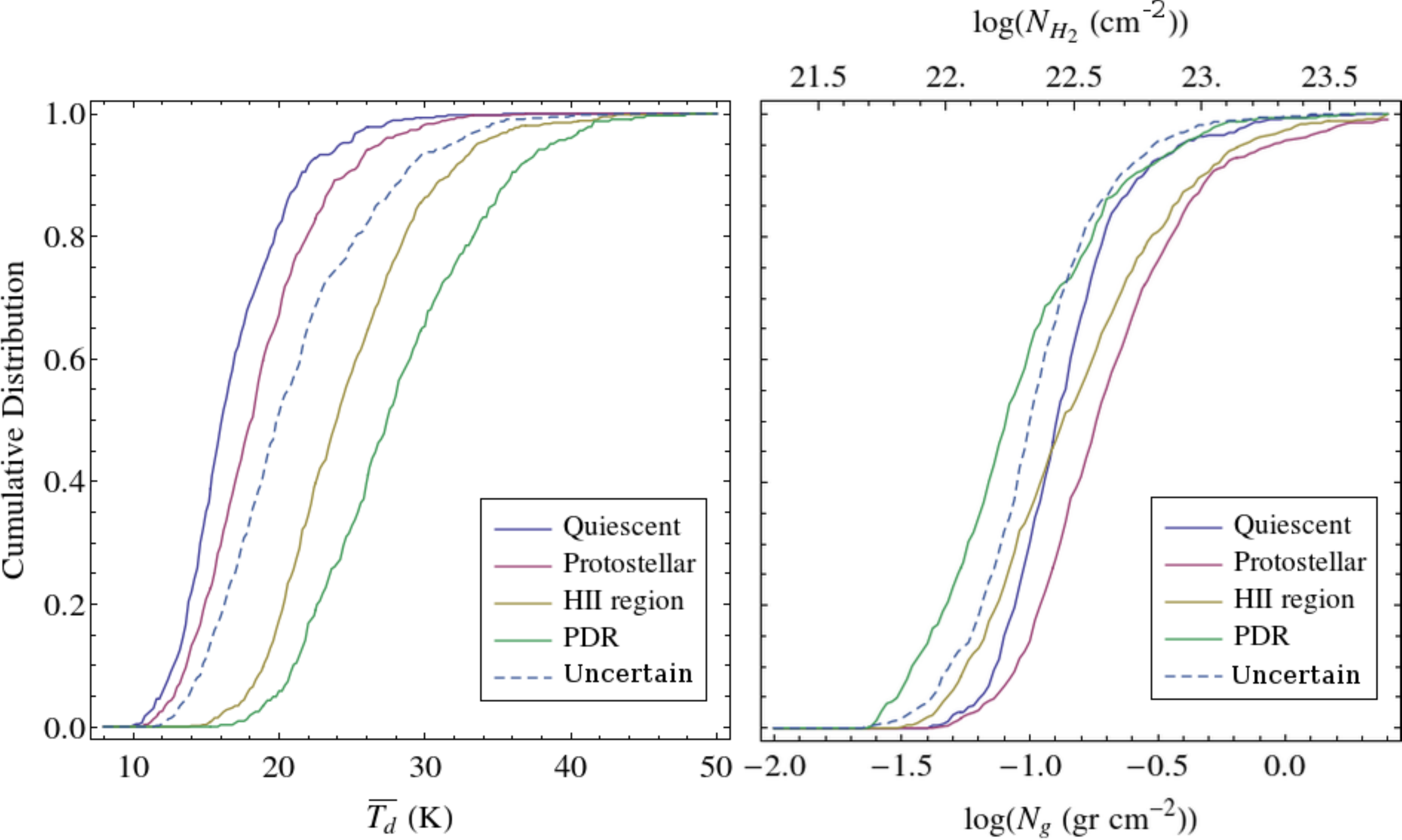} \figcaption{The
  empirical cumulative distribution functions of $\bar{T_d}$ (left panel)
  and $\log N_{g,{\rm P}}$ (right panel) for each evolutionary stage
  (continuous lines) and the Uncertain group (dashed lines).  The upper axis
  labeling of the right panel shows the equivalent logarithm of the  H$_2$ column density
  ($\log N_{\rm H_2}$).
\label{fig-cdf}}
\end{figure}

These differences are also apparent in Figure \ref{fig-cdf}, where solid
lines display the marginalized CDFs of
$\bar{T_d}$ and $\log N_{g,{\rm P}}$ for each evolutionary stage. The
dashed lines show the distributions of the Uncertain group. It is clear
from these plots that the median temperature increases monotonically with
evolutionary stage, and that the Protostellar and PDR clumps are the stages
associated with the largest and smallest column densities, respectively.
Figure \ref{fig-boxNT} shows Tukey box plots \citep[][Section
  5.9]{Feigelson2012msma} of the marginalized distributions of $\bar{T_d}$
and $\log N_{g,{\rm P}}$ separated by evolutionary stage.  In these plots,
the boxes indicate the interquartile range (half of the population), the
thick horizontal line inside each box indicates the median, and the error
bars encompass the data that are within 1.5 times the inter-quartile
distance from the box limits.  The remaining points, in all cases less than
4\% of the sample, are plotted individually with small circles and we refer
to them formally as outliers. Figure \ref{fig-boxNT} shows that the
$\bar{T_d}$ and $\log N_{g,{\rm P}}$ interquartile range shifts with
evolutionary stage. This is evidence of systematic differences between the
different populations, despite the large overlaps.  In practice, the
overlap between populations implies that it is unfeasible to construct
sensible predictive criteria that could determine the evolutionary stage of
a specific source based on its temperature and peak column density and it
also reflects that star formation is a continuous process that cannot be
precisely separated into distinct stages. Nevertheless, the fact that the
proposed evolutionary stages show a monotonic increase in mean temperature
demonstrates that the classification scheme has a legitimate physical
basis.

In the following, we focus our analysis on the Quiescent and Protostellar
populations.  Figures \ref{fig-sd} to \ref{fig-boxNT} show that these two
samples are similarly distributed and they exhibit the most important
overlap.  We test the statistical significance of the Quiescent and
Protostellar differences in $\bar{T_d}$ and $\log N_{g,{\rm P}}$ by
comparing these differences with their uncertainties.  Table
\ref{tab-means} shows the medians, means, and r.m.s. deviations of
$\bar{T_d}$, $T_{d,{\rm P}}$, $\log \bar{N_g}$, and $\log N_{g,{\rm P}}$
for each population. In general, these dispersions are larger than the
uncertainties of the individual values, indicating that the dispersions are
intrinsic to each population and not due to the fitting uncertainties.  The
means $\bar{T_d}$ for the Quiescent and Protostellar populations are $16.8$
and $18.6$ K, respectively.  The Protostellar population has a mean
$\bar{T_d}$ larger by $+1.8$ K compared with the Quiescent population.  We
can estimate the expected uncertainty of this mean difference using the
dispersion and size of each population, which gives
$\sqrt{\left(3.8^2/464\right)+\left(4.4^2/788\right)}\approx0.24$
K. Therefore, the difference is more than seven times the
expected uncertainty.  On the other hand, the difference of the means of
$\log N_{g,{\rm P}}$ is $+0.17$ in favor of the Protostellar
population. The expected uncertainty in this case is
$\sqrt{\left(0.25^2/464\right)+\left(0.34^2/788\right)}=0.017$, that is,
ten times smaller than the observed difference.  We conclude that the
observed differences of the Quiescent and Protostellar populations are
statistically significant.  Furthermore, the differences in temperature and
column density are orthogonal to the expected uncertainty correlation
(Figure \ref{fig-Dchi2con}), giving us more confidence that we are
observing a real effect in both parameters.  We confirm the significance of
the difference using more sophisticated statistical tests in Appendix
\ref{sec-stat}. Note that either the temperature or the column density
difference between other population pairs are larger than those between the
Quiescent and Protostellar.  

Are these statistical differences evident when comparing the $\bar{N_g}$
and $T_{d,{\rm P}}$?  On one hand, the difference between the mean
$T_{d,{\rm P}}$ of the Protostellar and Quiescent samples is 2.6 K, while
the expected uncertainty of this difference is 0.2 K. Therefore, despite
the larger fitting uncertainties of $T_{d,{\rm P}}$ (see Table
\ref{tab-NT}), we still detect a statistically significant difference
between both populations when comparing only their central temperatures.
On the other hand, the difference between the means of $\log \bar{N_g}$ is
$0.04$ over an expected uncertainty of $0.02$, that is, the difference is
only 2 times the uncertainty. The latter is not highly significant, which
is somehow expected because of the reasons explained at the beginning of
this section.  It is also expected from previous studies that indicate that
neither the mass of the clumps \citep{Hoq2013ApJ} nor their radii
\citep{Urquhart2014MNRAS} change conspicuously with evolutionary
stage. This in turn implies that the average column density should remain
approximately unchanged.

Within the Quiescent sample, we identified in Section \ref{sec-re70} a
population of 83 clumps that appear as far-IRDCs at 70 \um. Of these, there
are 77 associated with a single source along the line of sight.
This sample has a mean and
median $\bar{T_d}$ of $14.9$ and $14.7$ K, respectively. The remaining 
Quiescent population has mean and median $\bar{T_d}$ equal to
$17.2$ and $16.4$ K, respectively. 
Based on a Wilcoxon non-parametric test
\citep[Section 5.4]{Wall2012psa} { we obtain  a
  p-value of
$4\times10^{-7}$} under the null hypothesis that these distributions
are the same. {Therefore, the temperature differences between the far-IRDCs and the rest of the Quiescent clumps are
significant.}  The column densities of the far-IRDC subsample are
also larger compared to those of the rest of the
Quiescent sample. The far-IRDC $\log N_{g,{\rm P}}$ mean and median
are $-0.76$ and $-0.78$, respectively, while for the remaining 
Quiescent clumps they are $-0.88$ and $-0.91$, respectively.  Again,
we reject the null hypothesis (Wilcoxon p-value of
$\sim10^{-5}$) and  conclude that the far-IRDC sample is a colder and
denser subsample of the Quiescent population.
 
Finally, the Uncertain group (that is, MALT90 sources that could not be
classified into any evolutionary stage) seems to be a mixture of sources in
the four evolutionary classes, but associated with lower column densities
(median $\log N_{g,{\rm P}}\sim0.1$ gr cm$^{-2}$).  Figure \ref{fig-cdf}
shows that the $\bar{T_d}$ values of the Uncertain group distribute almost
exactly in between the other evolutionary stages.  Neither a Wilcoxon nor a
Kolmogorov-Smirnov test can distinguish the $\bar{T_d}$ distributions of
the Uncertain sample from the remainder of the MALT90 sources
combined with a significance better than 5\%.  Figure \ref{fig-cdf} also
shows that the column densities of the Uncertain group are in general lower
compared with those of any evolutionary stage except the PDRs.  Molecular
clumps with low peak column density may be more difficult to classify in
the mid-IR, since they are probably unrelated to high-mass star formation.
It is also possible that a significant fraction of these sources are
located behind the Galactic plane cirrus emission and possibly on the
far-side of the distance ambiguity, making the mid-IR classification more
difficult and decreasing the observed peak column density because of beam
dilution.

\subsubsection{Column Density and Temperature Evolution in Previous Studies}

Since the discovery of IRDCs \citep[typically dark at mid-IR
  wavelengths, see][]{Egan1998ApJ,Carey1998ApJ} it has been pointed
out that they  likely consist of cold  ($<20$ K) 
molecular gas. This  has been confirmed by several studies
of molecular gas  \citep{Pillai2006AA450,Sakai2008ApJ,Chira2013AA} and dust \citep[e.g.,][]{Rathborne2010ApJ}. 

Systematic \hh\ column density differences  between IR dark, quiescent,
and star-forming clumps have been more difficult to establish. Some
authors have found no significant column density differences between
these groups
\citep{Rathborne2010ApJ,Lopez-Sepulcre2010AA,Sanchez-Monge2013MNRAS}.
However, most studies based on large samples agree that star forming
clumps have larger molecular column densities compared to the quiescent ones
\citep{Dunham2011ApJ,Giannetti2013AA,Hoq2013ApJ,Csengeri2014AA,Urquhart2014MNRAS,He2015MNRAS}. 
Furthermore,
\citet{Beuther2002ApJ}, \citet{Williams2005AA}, and
\citet{Urquhart2014MNRAS} found evidence that  molecular clumps
which display star formation activity have a more concentrated density
profile.

\citet{Urquhart2014MNRAS}, based on ATLASGAL and the Red MSX Source
(RMS) survey \citep{Lumsden2013ApJ}, analyze a large ($\sim1300$)
number of molecular clumps with signs of high-mass star formation.
High-mass star formation activity was determined from associations with
the MSX point source catalog \citep{Egan2003VizieR}, methanol masers
\citep{Urquhart2013MNRAS431}, and \hii\ regions detected using
centimeter wavelength radio emission
\citep{Urquhart2007AA,Urquhart2009AA501,Urquhart2013MNRAS}.  In \citet{Urquhart2014MNRAS},  ATLASGAL
clumps associated with WISE sources \citep{Wright2010AJ} are
called massive star-forming (MSF) clumps and all the rest are otherwise
``quiescent.''
\citet{Urquhart2014MNRAS} find that MSF clumps have larger column
densities than their ``quiescent'' clumps by a factor of $\sim3$.

\citet{Urquhart2014MNRAS} and \citet{He2015MNRAS} 
also report that clumps associated with
\hii\ regions have larger column densities than the remainder of the star forming 
clumps.  This result contradicts our finding that \hii\ region sources
have typically lower column densities compared with the Protostellar
sample (see Table \ref{tab-means} and Figure \ref{fig-cdf}).  
To
examine this disagreement in more detail, we analyze the intersection
between the MSF and the MALT90 samples.  There are 515 MSF clumps in
common with the MALT90 sample that are covered by Hi-GAL: 285
classified as \hii\ regions, 204 as Protostellar, 22 as PDR, and 4 as
Quiescent. We calculate that these 515 sources have a mean average
temperature of 24 K and a mean log-peak column density of $-0.63$. The
temperature is consistent with the \hii\ region sample of MALT90, but
the column densities are much higher.  Within these 515 sources 
 we find that, in agreement with \citet{Urquhart2014MNRAS},
those with centimeter wavelength emission have significantly higher
column densities ($\log N_{g,{\rm P}}=-0.59$) and temperatures ($26$
K) compared with the rest ($\log N_{g,{\rm P}}=-0.67$ and $\bar{T_d}=22$
K).

The reason for \citet{Urquhart2014MNRAS} finding that sources associated
with \hii\ regions are associated with the largest column densities, in
disagreement with our results, arises most likely from differences in the
classification criteria.  \citet{Urquhart2014MNRAS} report centimeter radio
emission arising from ionized gas toward 45 out of the 204 common sources
we classify as Protostellar, and 94 out of the 285 clumps we classify as
\hii\ regions were observed by the CORNISH survey at 5 GHz \citep[$\sim2$
  mJy sensitivity,][]{Hoare2012PASP} and were not detected.  These are
relatively few sources and exchanging their classification (Protostellar by
\hii\ region and vice-versa) does not modify the trends described in the previous
section.  However, if they reflect an underlying fraction of misclassified
sources between the Protostellar and \hii\ region groups, they might change
the statistics.

Conversely, we detect embedded HMYSOs in 641 ATLASGAL
sources that are treated as ``quiescent'' in \citet{Urquhart2014MNRAS}, in
part due to the better sensitivity and angular resolution of MIPS compared
to MSX and WISE. It is likely that the ``quiescent'' sample of
\citet{Urquhart2014MNRAS} does not contain currently young high-mass stars,
but does contain a large fraction of intermediate mass star formation
activity, and some of these sources are also associated with PDRs.  In
summary, we expect that the Quiescent sample from MALT90 to be more truly
devoid of star formation than the non-MSF ATLASGAL clumps, while at the same
time, several of our Protostellar clumps are probably associated with
\hii\ regions, which are more efficiently detected using radio centimeter
observations.


{\subsubsection{Temperature and Column Density Contrasts}\label{sec-cont}}

We analyze spatial variations of $T_d$ and $N_g$ by comparing their
values at the peak intensity position with the average value in the
clump. For each MALT90 clump, we define the temperature contrast and
log-column density contrast as $\Delta T=\bar{T_d}-T_{d,{\rm P}}$ and
$\Delta\log N_g=\log\left(\bar{N_g}/N_{g,{\rm P}}\right)$,
respectively.

Table \ref{tab-cont} {lists} the means and medians of the temperature and
log-column density contrasts. Table \ref{tab-cont} also {gives} 95\%
confidence intervals\footnote{The upper limit of the CI is the lowest value
  $u$ larger than the observed  median for which we can reject the null
  hypothesis that $u$ is the true population median with a significance of
  5\%. The lower limit of the CI is calculated similarly.} (CIs) for the  medians of $\Delta T$ and
$\Delta\log N_g$ per evolutionary stage, determined using the sign test
\citep[Section 12.2]{Ross2004ipses}. They were calculated using the task
\texttt{SIGN.test} from the R statistical suite\footnote{www.r-project.org}
(version 3.1.1).  The sign test
is not very sensitive but it has the advantage that it is non-parametric
and, in contrast to the Wilcoxon test (for example), it does not assume that
the distributions have the same shape.

A negative $\Delta\log N_g$ indicates that the clump has a centrally peaked
column density profile with the absolute value of $\Delta\log N_g$ being a
measure of its steepness. As a reference, a critical Bonnor-Ebert sphere is
characterized by $\Delta\log N_g=-0.5515$.  Perhaps not very surprisingly,
the $\Delta\log N_g$ means, medians, and CIs are always negative,
indicating that most of the clumps are centrally peaked.  We  find that
the medians are significatively different between evolutionary stages, with
no overlap in the CIs.  The \hii\ region clumps are  those
associated with the steepest column density profiles, followed by the Protostellar and the PDR clumps. Clumps in the Quiescent evolutionary stage  are associated with 
the smoothest column density profiles.

The temperature contrasts are also distinct for different evolutionary
stages.  A positive $\Delta T$ indicates that the dust 
 temperature increases away from the clump center, that is, dust temperature at the peak column density
  position ($T_{d,{\rm P}}$) being lower than the average temperature
  ($\bar{T_d}$).  On the other hand, $\Delta T$ is negative for decreasing
  temperature profiles.  $\Delta T$ is positive for Quiescent clumps and
PDRs, it is consistent with zero for the Protostellar sources (temperature
at peak similar to average temperature), and positive (peak column density
warmer than average temperature) for the \hii\ region sample.

\subsection{\boldmath Mid-IR Classification versus $T_d$, $N_g$, $\Delta T_d$, and $\Delta\log N_g$}

\begin{figure}
\includegraphics[width=\textwidth]{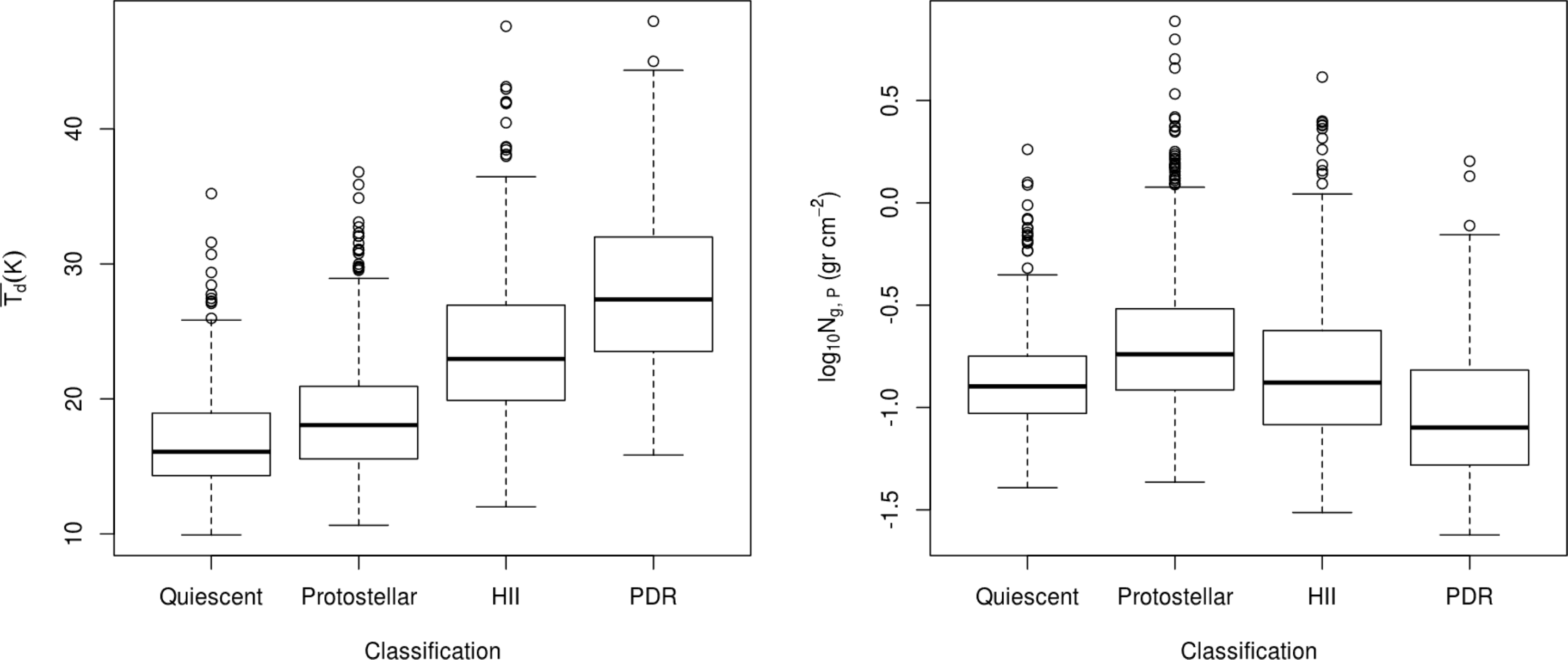}
\figcaption{Box-plots of the marginalized temperature and log-peak
  column density distributions. The thick line indicates the median,
  the boxes enclose the interquartile range (50\% of the total
  population), and the error bars indicate the minimum and maximum
  data points that are within $1.5$ times the inter-quartile distance
  from the boxes' limits. The remaining points are individually
  marked with circles as outliers. \label{fig-boxNT}}
\end{figure}

The previous sections have presented the differences between the
temperature and column densities of the MALT90 groups.  These differences
are qualitatively consistent  with the evolutionary sequence sketched in
\citet{Jackson2013PASA} that starts with the Quiescent, and proceeds
through the Protostellar, \hii\ region, and PDR evolutionary stages.  As Figure \ref{fig-boxNT}
shows, Quiescent clumps are the coldest, in agreement  with the
expectation that these clumps are starless and there are no embedded
young high-mass stars.  The  far-IRDC subsample
of the Quiescent population is colder and denser on average compared
to the rest of the Quiescent clumps, and they might represent a late
pre-stellar phase just before the onset of star-formation. The mean
temperature and $\log N_{g,{\rm P}}$ of the far-IRDC subsample are
$\sim15$ K and $-0.78$, respectively.

To establish what fraction of the Quiescent clumps might evolve to form
high-mass stars, we use for now criteria defined by previous authors based
on distance independent information, such as the column density; a more
complete analysis will be done in Contreras et al.\ (in preparation).
\citet{Lada2010ApJ} and \citet{Heiderman2010ApJ} propose that the star
formation rate in a molecular cloud is proportional to the mass of gas
with column densities in excess of $\sim$120~\Msun~pc$^{-2}$
($\sim2.43\times10^{-2}$ gr cm$^{-2}$).  Since there is considerable
overlap in column density between MALT90 clumps that have different levels
of star formation activity, we start by assuming that this
relation gives the average star formation rate over the timescale of 2 Myr
adopted by \citet{Lada2010ApJ} and \citet{Heiderman2010ApJ}.  We find that
98\% of the Quiescent clumps have $\bar{N_g}>120$~\Msun~pc$^{-2}$, including
all of the far-IRDCs, which suggest that most of these clumps will support
some level of star formation activity in the future.  
\citet{Urquhart2014MNRAS} propose a column density threshold of $0.05$~gr~cm$^{-2}$ for what
they denominate ``effective'' high-mass star formation. 
This same threshold 
 was recently proposed by \citet{He2015MNRAS} based on a study of $405$ ATLASGAL sources.
Of the Quiescent sample, 78\% of the
clumps have an average column density above this threshold, with the percentage increasing to 92\% for the far-IRDCs. 
Based on these  criteria, we conclude that
virtually all Quiescent clumps will develop at least low-mass star
formation activity and that a large fraction ($>70\%$) will form high-mass
stars. 
%
On the other hand, \citet{Lopez-Sepulcre2010AA} suggest a third column density threshold
based on the observed increase of molecular outflows for clumps with column
densities in excess of $0.3$ gr cm$^{-2}$.  This column density is
significatively larger than the previous thresholds, and only 3\% and 6\%
of the Quiescent and far-IRDCs populations, respectively, have larger average
column densities.  However, half of the clump sample of
\citet{Lopez-Sepulcre2010AA} have diameters $< 35\arcsec$ (the beam size of
our column density maps) and more than a third have masses $< 200 \Msun$,
which indicates that the $0.3$ gr cm$^{-2}$ threshold may be pertinent
for more compact structures than the clumps considered in this work.

The temperature and temperature contrast of the Quiescent clumps are
qualitatively consistent with equilibrium between the interstellar
radiation field and dust and gas cooling \citep{Bergin2007ARA&A}.  We find
that Quiescent clumps are the coldest among the evolutionary stages, but
they are typically warmer ($\sim17$ K) than expected from thermal
equilibrium between dust cooling and cosmic ray heating alone ($T_d\sim10$
K).  We also find that the central regions of the Quiescent clumps are
in general colder than their external layers ($\Delta T$ negative). 
These characteristics  are consistent with Quiescent clumps  being  
heated by a combination of external radiation and cosmic-rays.
The Quiescent sources also have
the flattest density structure, with the {largest}
 $\Delta \log N_g$ among all
the other evolutionary stages. This is similar to  the behavior found
by \citet{Beuther2002ApJ}, that is, the earliest stages of
high-mass star formation are characterized by flat density profiles that
become steeper as they collapse and star formation ensues.

The Protostellar clump sample can be distinguished from Quiescent clumps
based on their column density and dust temperature.  Protostellar clumps
have larger column densities ($\sim0.2$ gr cm$^{-2}$) and are slightly
warmer ($\sim19$~K). The central temperatures of the Protostellar clumps
also increase and become comparable to the temperature in their outer
regions ($\Delta T\cong0$).  These characteristics indicate that
Protostellar clumps have an internal energy source provided by the
HMYSOs. According to the results presented by \citet{Hoq2013ApJ}, there is
no significative difference in the distribution of masses between the
Quiescent and Protostellar population. If we assume that this is also the
case for the sample presented in this work \citep[which will be confirmed in
upcoming publications, see also][]{He2015MNRAS}, then the most likely reason for the larger column
densities of the Protostellar sample compared with the Quiescent sample is
gravitational contraction.  Because contraction develops faster on the
densest, central regions, we expect the column density profiles to become
steeper at the center of the clump. This is consistent with the observed
decrease of $\Delta \log N_g$ for the Protostellar clumps compared with
the Quiescent clumps.


The \hii\ region sample is associated with the most negative temperature and
column density contrasts {(median of $\Delta T=-0.33$ K and $\Delta \log
N_g=-0.42$)} compared with any other population, which indicates that
\hii\ region clumps are very concentrated and they have a strong
central heating source.  This picture is consistent with the presence
of a young high-mass star in the center of the clump.  The slight
decrease of the peak column density compared with the Protostellar
phase {could} be explained because of the expansion induced by the
development of the \hii\ region and the fraction of gas mass that has
been locked into newly formed stars.

Finally, PDR clumps have the lowest column densities and
largest temperatures among the four evolutionary stages.  They are also
associated with having colder temperatures toward the center compared to their
outer regions.  PDR clumps are possibly the remnants of molecular clumps
that have already been disrupted by the high-mass stars' winds, strong UV radiation field, and the expansion of 
\hii\ regions.  These molecular remnants are being illuminated and heated
from the outside by the newly formed stellar population, but probably  are
neither dense nor massive enough to be 
able to sustain further high-mass star formation.

{\section{SUMMARY}\label{sec-sum}}

We determined dust temperature and column density maps toward
\WithFit\ molecular clumps. This number corresponds to more than 99\% of
the ATLASGAL sources that form the MALT90 sample. We fit greybody
models to far-IR images taken at 160, 250, 350, 500, and 870 \um.  This
catalog represents the largest sample of high-mass clumps for
which both dust temperature and column density have been simultaneously
estimated.  We summarize the main results and conclusions as follows.
\begin{enumerate}
\item{The average dust temperature increases monotonically along the
  proposed evolutionary sequence, with median temperatures ranging from $16.1$ K
  for the Quiescent clumps to $27.4$ K for the clumps associated with
  PDRs. This confirms that the MALT90 mid-IR classification broadly
  captures the physical state of the molecular clumps.}
\item{The highest column densities are associated with the
  Protostellar clumps, that is, those that show mid-IR signs of star
  formation activity preceding the development of an
  \hii\ region. The average peak column density of the Protostellar
  clumps is 0.2~gr~cm$^{-2}$, which is about 50\% higher than the peak
  column densities of clumps in the other evolutionary
  stages. We interpret this as evidence of gravitational contraction
  or  possibly that Protostellar clumps are more massive.
  The latter possibility will be analyzed in future work (Contreras et
  al., in preparation).}
\item{The radial temperature gradients within the clumps decrease from
  positive (higher temperatures in the outer layers of the clump), to null
  (no dust temperature gradient), and to negative (higher temperatures
  toward the center of the clump) values associated with the Quiescent,
  Protostellar, and \hii\ region clumps, respectively.  Quantitatively,
  the mean difference between the average ($\bar{T_d}$) and the
    central ($T_{d,{\rm P}}$) clump temperatures range
    between $+0.7$, $-0.1$, and $-0.6$ K for the Quiescent, Protostellar,
    and \hii\ region samples, respectively. This confirms that Quiescent
  clumps are being externally heated and Protostellar and \hii\ region
  clumps have an internal embedded energy source.}
\item{The ratio between the peak and average column density for each clump
  category ranges between 1.8 and 2.6.  The flattest column density
  profiles are associated with the Quiescent population, becoming steeper
  for the Protostellar {and} \hii\ region clumps. This is qualitatively
  consistent with the hypothesis of evolution through gravitational
  contraction, in which the contrast is a measure of evolutionary progress.}
\item{The PDR clump population is characterized by low column densities
  ($\sim0.09$~gr~cm$^{-2}$), high temperatures ($27$ K), and
  a positive radial temperature gradient (colder inner regions toward warmer
  dust on the outside).  We interpret this as evidence that these
  sources are the externally illuminated remnants of molecular clumps already
  disrupted by high-mass star formation feedback.}
\item{We identify $83$ far-IR dark clouds, that is, Quiescent clumps
  that appear in absorption at 70 \um\ against the Galactic
  background. These clumps are cooler and they have  higher column
  densities compared to the remainder of the Quiescent population. Therefore, 
   they are likely in the latest stage of pre-stellar
  contraction or they  may represent a more massive subsample of the Quiescent clumps.}
\end{enumerate}

\acknowledgements{A.E.G. and H.A.S.  acknowledges support from NASA Grants
  NNX12AI55G and NNX10AD68G.  A.E.G.  acknowledge partial support from
  CONICYT through project PFB-06 and FONDECYT grant 3150570.
  J.M.J. acknowledges support from NASA Grant NNX12AE42G and NSF grant
  AST-1211844. We thank G.\ Garay and an anonymous referee for careful
  reading and helpful comments.}

\appendix


{\section{Significance Tests on the Difference between the Quiescent and Protostellar Populations}\label{sec-stat}}

We analyze the statistical significance of the difference between the
Quiescent and Protostellar populations using two non-parametric tests: the
sign test \citep[Section 12.2]{Ross2004ipses} and the Kolmogorov-Smirnov
(K-S) two sample test \citep[Section 5.4]{Wall2012psa}. We implement them
using the R statistical suite (version 3.1.1) through the tasks
\texttt{SIGN.test} and \texttt{ks.test}.

We apply the sign test to evaluate whether the medians of the $\bar{T_d}$
and $\log{N_{g,{\rm P}}}$ distributions of the Quiescent and Protostellar
samples are significativelly different.  Because these two
samples are the most similarly distributed, 
we can  conclude that the four evolutionary stages can be distinguish based on their $\bar{T_d}$
or $\log{N_{g,{\rm P}}}$ distributions if we can show it for the Quiescent and Protostellar
samples.

The median 95\% CIs of $\bar{T_d}$
and of $\log{N_{g,{\rm P}}}$ are given in Table \ref{tab-sign}, which shows
that the CIs of the Quiescent and Protostellar populations do not overlap.
This means that there is no value that is {simultaneously} consistent (at 5\% significance
level)\footnote{A value $m$ being consistent with the median of population
  $P$ at 5\% confidence level means that we cannot reject the null
  hypothesis that the median of $P$ is $m$ at a 5\% significance level
  using the sign test.}  with it being the median of the Quiescent and of
the Protostellar populations.  We call the Original sample (first row of
Table \ref{tab-sign}) that of MALT90 sources analyzed by default throughout
this work, that is, the Quiescent and Protostellar sources which are not
blended along the same line of sight with another MALT90 source, and with
the Quiescent sources associated with compact 70 \um\ emission
re-classified as Protostellar (see Section \ref{sec-re70}).  The second row
of Table \ref{tab-sign} gives the CIs without applying this
re-classification based on the 70 \um\ images. The third row shows the CIs
associated with the samples from where we have removed the outliers
displayed in Figure \ref{fig-boxNT}. Finally, the sample of the fourth row
takes all MALT90 sources in each category regardless of being multiple
sources across the same line of sight.  In all cases, we see that the CIs
do not overlap.  We conclude that the $\bar{T_d}$ or $\log{N_{g,{\rm P}}}$
medians are significantly different between the samples and that these
differences do not depend critically on the censoring or re-classifications
applied in this work.

Table \ref{tab-sigs} shows the results of the K-S test p-values associated
with the null hypothesis that the distributions are the same.  
We conclude that the probability of the observed data 
  assuming that both samples are drawn from the same distribution is $<10^{-10}$ in
  all cases.  Therefore, we reject the null hypothesis in each case, and
conclude that the statistical evidence does not support that the Quiescent
and Protostellar samples come from the same underlying distribution. We
conclude that the differences between these two evolutionary stages are
significant and robust, i.e., they are independent of possibly
misclassified sources or outliers.


\bibliographystyle{apj}
\bibliography{bibliografia}

\begin{deluxetable}{lccccl}
    \tablewidth{0pc} \tablecolumns{6} \tabletypesize{}
    \tablecaption{Observational Characteristics of the Data\label{tab-ins}}
    \tablehead{   \colhead{Instrument} & \colhead{Band} & \colhead{FWHM\tablenotemark{a}} & \colhead{$\sigma_p$} & \colhead{$\sigma$} &\colhead{Telescope}\\ \colhead{~} & \colhead{(\um)} & \colhead{(\arcsec)} & \colhead{(mJy)} & \colhead{(MJy~sr$^{-1}$)}&\colhead{~} }
 \startdata
PACS  &  70 &  9.2 & 18  & 1.1   & HSO \\
PACS  & 160 & 12.0 & 24  & 1.5   & HSO \\
SPIRE & 250 & 17.0 & 12  & 0.7   & HSO \\
SPIRE & 350 & 24.0 & 12  & 0.7   & HSO \\
SPIRE & 500 & 35.0 & 12  & 0.7   & HSO \\
LABOCA& 870 & 19.2 & 25  & 1.5   & APEX \\
 \enddata
\tablecomments{The noise displayed for the SPIRE  data is taken in nominal mode. Noise associated with data taken in bright mode is $\sim$4 times higher.}
 \tablenotetext{a}{The beam FWHM associated with each waveband of the PACS and SPIRE data was adopted from \cite{Olmi2013AA}.}
%
\end{deluxetable}%

\begin{deluxetable}{llcc}
    \tablewidth{0pc} \tablecolumns{4} \tabletypesize{}
    \tablecaption{HSO  Observations \label{tab-ids}}
    \tablehead{ \colhead{Target} & \colhead{Obs.\ ID} & \colhead{Obs.\ Mode\tablenotemark{a}} & \colhead{Obs.\ Date}}
\startdata
Field-000\_0 & 1342204102/3& PB & 2010-09-07  \\
Field-002\_0 & 1342204104/5& PB & 2010-09-07  \\
Field-004\_0 & 1342214761/2& PB & 2011-02-24  \\
Field-006\_0 & 1342214763/4& PN & 2011-02-24  \\
Field-008\_0 & 1342218963/4& PN & 2011-04-09  \\
Field-011\_0 & 1342218965/6& PN & 2011-04-09  \\
Field-013\_0 & 1342218999/0& PN & 2011-04-10  \\
Field-015\_0 & 1342218997/8& PN & 2011-04-10  \\
Field-017\_0 & 1342218995/6& PN & 2011-04-10  \\
Field-019\_0 & 1342218644/5& PN & 2011-04-15  \\
Field-022\_0 & 1342218642/3& PN & 2011-04-15  \\
Field-283\_0 & 1342255009/0& PN & 2012-11-14  \\
Field-286\_0 & 1342255011/060& PN & 2012-11-14  \\
Field-288\_0 & 1342255061/2& PN & 2012-11-15  \\
Field-299\_0 & 1342183075/6& PN & 2009-09-03  \\
Field-301\_0 & 1342203083/4& PN & 2010-08-15  \\
Field-303\_0 & 1342189081/2& PN & 2010-01-08  \\
Field-305\_0 & 1342189083/4& PN & 2010-01-08  \\
Field-308\_0 & 1342203085/6& PN & 2010-08-16  \\
Field-310\_0 & 1342203278/9& PN & 2010-08-20  \\
Field-312\_0 & 1342189109/0& PN & 2010-01-09  \\
Field-314\_0 & 1342203280/1& PN & 2010-08-21  \\
Field-316\_0 & 1342203282/3& PN & 2010-08-21  \\
Field-319\_0 & 1342203289/0& PN & 2010-08-21  \\
Field-321\_0 & 1342203291/2& PN & 2010-08-21  \\
Field-323\_0 & 1342189878/9& PN & 2010-01-29  \\
Field-325\_0 & 1342203293/4& PN & 2010-08-22  \\
Field-327\_0 & 1342204042/3& PN & 2010-09-03  \\
Field-330\_0 & 1342204044/5& PN & 2010-09-04  \\
Field-332\_0 & 1342204046/7& PN & 2010-09-04  \\
Field-334\_0 & 1342204054/5& PN & 2010-09-04  \\
Field-336\_0 & 1342204056/7& PN & 2010-09-05  \\
Field-338\_0 & 1342204058/9& PN & 2010-09-05  \\
Field-341\_0 & 1342204094/5& PN & 2010-09-06  \\
Field-343\_0 & 1342204092/3& PN & 2010-09-06  \\
Field-345\_0 & 1342204090/1& PN & 2010-09-06  \\
Field-347\_0 & 1342204100/1& PN & 2010-09-06  \\
Field-349\_0 & 1342214510/1& PN & 2011-02-20  \\
Field-352\_0 & 1342214575/6& PN & 2011-02-20  \\
Field-354\_0 & 1342214713/4& PN & 2011-02-24  \\
Field-356\_0 & 1342204368/9& PB & 2010-09-12  \\
Field-358\_0 & 1342204366/7& PB & 2010-09-12  \\
NGC6334-S & 1342239908& SB & 2012-03-01  \\
NGC6334-N & 1342239909& SB & 2012-03-01  \\
M17-S & 1342241160& SB & 2012-03-04  \\
\enddata
\tablenotetext{a}{PN: Parallel Normal mode, PB: Parallel Bright mode, SB: SPIRE Bright mode.}
\end{deluxetable}
\begin{deluxetable}{lcccccccccccccl}
   \rotate \tablewidth{0pc} \tablecolumns{15} \tabletypesize{\scriptsize}
   \tablecaption{Temperature and Dust Column Density of MALT90
     Sources\label{tab-NT}} 
   \tablehead{\colhead{Source\tablenotemark{a}} &\colhead{Effective} &\multicolumn{3}{c}{Average} & 
     \multicolumn{3}{c}{Temperature} &
     \multicolumn{3}{c}{Average Column}&
     \multicolumn{3}{c}{Peak Column} &\colhead{Mid-IR}\\ 
     \colhead{}&\colhead{Radius}&\multicolumn{3}{c}{Temperature} &
     \multicolumn{3}{c}{at Peak}&
     \multicolumn{3}{c}{Density}&
     \multicolumn{3}{c}{Density}&\colhead{Classification\tablenotemark{b}}\\ 
     \colhead{} &\colhead{(\arcsec)}& \multicolumn{3}{c}{(K)} & \multicolumn{3}{c}{(K)}&
     \multicolumn{3}{c}{(gr cm$^{-2}$)} & \multicolumn{3}{c}{(gr cm$^{-2}$)}&\colhead{}\\
   \colhead{}&\colhead{}&\colhead{$\bar{T_d}$}&\colhead{$\sigma^-$}&\colhead{$\sigma^+$}&\colhead{$T_{d,{\rm P}}$}&\colhead{$\sigma^-$}&\colhead{$\sigma^+$}&\colhead{$\bar{N_g}$}&\colhead{$\log \bar{N_g}$}&\colhead{$\sigma$}&\colhead{$N_{g,{\rm P}}$}&\colhead{$\log N_{g,{\rm P}}$}&\colhead{$\sigma$}&\colhead{}\\
\colhead{(1)}&\colhead{(2)}&\colhead{(3)}&\colhead{(4)}&\colhead{(5)}&\colhead{(6)}&\colhead{(7)}&\colhead{(8)}&\colhead{(9)}&\colhead{(10)}&\colhead{(11)}&\colhead{(12)}&\colhead{(13)}&\colhead{(14)}&\colhead{(15)}}
   \startdata
   \multicolumn{15}{c}{\nodata}\\
AGAL000.254$-$00.496 & 26.2 & 19.3 & 0.4 & 0.4 & 20. & 1. & 1. & 0.052 & $-$1.28 & 0.02 & 0.074 & $-$1.13 & 0.04 & Quiescent$^C$ \\
AGAL000.259$+$00.017 & 87.6 & 20.8 & 0.1 & 0.1 & 19. & 1. & 1. & 0.506 & $-$0.296 & 0.004 & 0.977 & $-$0.01 & 0.03 & Quiescent \\
AGAL000.264$+$00.032 & 43.0 & 20.5 & 0.3 & 0.3 & 19. & 1. & 1. & 0.493 & $-$0.307 & 0.009 & 0.832 & $-$0.08 & 0.03 & Quiescent \\
AGAL000.268$-$00.084$^{\rm m}$ &  44.5  &  25.2  &  0.4  &  0.4  &  23.  &  2.  &  2.  &  0.056  &  $-$1.25  &  0.01  &  0.115  &  $-$0.94  &  0.04  &  Quiescent$^D$ \\
AGAL000.273$-$00.064$^{\rm m}$ &  59.2  &  26.7  &  0.3  &  0.4  &  25.  &  2.  &  2.  &  0.093  &  $-$1.031  &  0.009  &  0.195  &  $-$0.71  &  0.04  &  Uncertain \\
AGAL000.274$-$00.501 & 36.3 & 25.9 & 0.5 & 0.6 & 27. & 3. & 3. & 0.037 & $-$1.43 & 0.02 & 0.06 & $-$1.22 & 0.05 & PDR \\
AGAL000.281$-$00.482 & 58.3 & 21.2 & 0.2 & 0.2 & 27. & 3. & 3. & 0.093 & $-$1.03 & 0.01 & 0.186 & $-$0.73 & 0.05 & \hii\ region \\
   \multicolumn{15}{c}{\nodata}\\
   \enddata
   \tablecomments{Parameters and uncertainties are given by $\bar{T_{\!d}}^{\,\sigma^+}_{\,\sigma_-}$ and $\log(N_g)\pm\sigma$.\\The logarithm of the mass column density is related to the logarithm of the number column density of molecules and \hh\ through: \mbox{$\log N_g(\text{gr cm$^{-2}$})=-23.4146+\log N_p(\text{cm$^{-2}$})=-23.3291+\log N_{\rm H_2}(\text{cm$^{-2}$})$}.}
\tablenotetext{a}{The ``m'' superscript indicates that in the direction of this ATLASGAL clump there are multiple MALT90 sources (Rathborne et al.\ in preparation).}
\tablenotetext{b}{The ``C'' and  ``D'' superscripts indicate 
  Quiescent clumps with a 70
  \um\ compact emission  source or far-IRDCs, respectively. No subscript means
  that we do not detect either of these morphological features at 70
  \um\ (see Section \ref{sec-re70}). }
\end{deluxetable}
\begin{deluxetable}{lccccccccccccc}
   \rotate
 \tablewidth{0pc} \tablecolumns{14} \tabletypesize{\small}
   \tablecaption{Average and Dispersion of Dust Temperature and Log-Column
     Density Distributions\label{tab-means}}
\tablehead{
\colhead{Classification} &\colhead{Number}&\multicolumn{3}{c}{$\bar{T_d}$}&\multicolumn{3}{c}{$T_{d,{\rm P}}$}&\multicolumn{3}{c}{$\log \bar{N_{g}}$}&\multicolumn{3}{c}{$\log N_{g,{\rm P}}$}\\
\colhead{}&\colhead{}&\colhead{Median}&\colhead{Mean}&\colhead{Dev.} &\colhead{Median}&\colhead{Mean}&\colhead{Dev.}&\colhead{Median}&\colhead{Mean}&\colhead{Dev.} &\colhead{Median}&\colhead{Mean}&\colhead{Dev.}\\
\colhead{}&\colhead{}&\multicolumn{6}{c}{(K)} & \multicolumn{6}{c}{$\log(\textrm{gr~cm}^{-2})$}
}
\startdata
Quiescent    & 464 & 16.1 & 16.8 & 3.8 &  15.6 & 16.1 & 3.4 &  $-1.11 $ & $-1.11$ & 0.27 &  $-0.90 $ & $-0.86$ &  0.25\\
Protostellar & 788 & 18.1 & 18.6 & 4.4 &  18.1 & 18.7 & 4.4 &  $-1.10 $ & $-1.07$ & 0.31 &  $-0.74 $ & $-0.69$ &  0.34\\
\hii\ region & 767 & 23.0 & 23.7 & 5.2 &  23.6 & 24.3 & 5.4 &  $-1.29 $ & $-1.24$ & 0.33 &  $-0.88 $ & $-0.82$ &  0.35\\
PDR  &         326 & 27.4 & 28.1 & 5.9 &  26.4 & 27.3 & 5.8 &  $-1.44 $ & $-1.40$ & 0.30 &  $-1.10 $ & $-1.04$ &  0.34\\
Uncertain &    562 & 19.8 & 20.9 & 5.4 &  19.5 & 20.2 & 5.1 &  $-1.26 $ & $-1.26$ & 0.30 &  $-1.00 $ & $-0.98$ &  0.27\\
\enddata
\tablecomments{Dev.\ represents root mean square deviation.}
\end{deluxetable}
\begin{deluxetable}{lccrccr}
 \tablewidth{0pc} \tablecolumns{7} \tabletypesize{}
   \tablecaption{Temperature and Log-Column Density Contrasts\label{tab-cont}}
\tablehead{
\colhead{Classification} & \multicolumn{3}{c}{$\Delta T$}&\multicolumn{3}{c}{$\Delta\log N_g$}\\
\colhead{}& \colhead{Mean}& \colhead{Median}&\colhead{95\% CI}&\colhead{Mean}& \colhead{Median}&\colhead{95\% CI}
}
\startdata
Quiescent    & \phs$0.69$  & \phs$0.46$  & $[$\phs$0.39 ,$\phs$ 0.54]$ & $-0.24$ & $-0.23$ &$[-0.24,-0.22]$\\
Protostellar &    $-0.09$  &     $0.04$  & $[     -0.03 ,$\phs$ 0.10]$ & $-0.38$ & $-0.37$ &$[-0.39,-0.36]$\\
\hii\ region &    $-0.58$  &    $-0.33$  & $[     -0.43 ,      -0.22]$ & $-0.42$ & $-0.42$ &$[-0.43,-0.40]$\\
PDR          & \phs$0.79$  & \phs$0.54$  & $[$\phs$0.40 ,$\phs$ 0.75]$ & $-0.36$ & $-0.34$ &$[-0.37,-0.33]$\\
\enddata
\tablecomments{$\Delta\log N_g=\log\left(\bar{N_g}/N_{g,{\rm P}}\right)$ and $\Delta T=\bar{T_d}-T_{d,{\rm P}}$ (Section \ref{sec-cont}).  For values inside  the CIs we cannot reject the hypothesis that they are the  median  with a significance better than 5\% using the sign test. }
\end{deluxetable}

\begin{deluxetable}{lcccc}
    \tablewidth{0pc} \tablecolumns{5} \tabletypesize{}
    \tablecaption{Robustness of the Quiescent/Protostellar Differences: Sign Test\label{tab-sign}}
\tablehead{
\colhead{Sample}& \multicolumn{2}{c}{Temperature 95\% CI (K)     } & \multicolumn{2}{c}{$\log{(N_{g,{\rm P}})} $ 95\% CI}\\
\colhead{} & \colhead{Quiescent} &\colhead{Protostellar}& \colhead{Quiescent} &\colhead{Protostellar}
}
\startdata
Original                     & $[15.7, 16.4]$  & $[17.6, 18.4]$ & $[-0.91, -0.86]$& $[-0.76, -0.72]$\\
No 70 \um\ re-classification & $[16.0, 16.6]$  & $[17.6, 18.4]$ & $[-0.91, -0.86]$& $[-0.74, -0.69]$\\
No outliers                  & $[15.6, 16.3]$  & $[17.5, 18.3]$ & $[-0.92, -0.88]$& $[-0.77, -0.73]$\\
With multiple sources        & $[16.0, 16.7]$  & $[17.7, 18.4]$ & $[-0.91, -0.87]$& $[-0.76, -0.72]$\\
\enddata
\end{deluxetable}

\begin{deluxetable}{lcc}
    \tablewidth{0pc} \tablecolumns{3} \tabletypesize{}
    \tablecaption{Robustness of the Quiescent/Protostellar Differences: K-S test\label{tab-sigs}}
\tablehead{
\colhead{}& \multicolumn{2}{c}{ p-values} \\
   \colhead{Sample} & \colhead{$\bar{T_d}$} &\colhead{$\log{N_{g,{\rm P}}} $} }
\startdata
Original                     & $4\times10^{-12}$ & $<2.2\times10^{-16}$ \\  
No 70 \um\ re-classification & $2\times10^{-10}$ & $<2.2\times10^{-16}$ \\   
No outliers                  & $3\times10^{-12}$ & $<2.2\times10^{-16}$ \\    
With multiple sources        & $5\times10^{-11}$  & $<2.2\times10^{-16}$ \\    
\enddata
\end{deluxetable}
\clearpage



\end{document}